\documentclass[12pt]{article}
\usepackage{amssymb,amsmath}
\usepackage{bm} 
\usepackage{booktabs} 
\usepackage{array}
\usepackage{latexsym}
\usepackage{graphicx}
\usepackage{color}
\usepackage{slashed}
\usepackage{datetime}
\usepackage[nosort]{cite}
\usepackage{verbatim}
\usepackage{enumerate}
\usepackage{makecell} 
\usepackage{chngpage} 
\allowdisplaybreaks
\usepackage{psfrag}
\usepackage{mathtools}
\usepackage{mciteplus}
\usepackage{dsfont}
\usepackage{multirow} 

\usepackage[colorlinks=true,      linkcolor=blue,      urlcolor=blue,      
            filecolor=blue,      citecolor=blue,       pdfstartview=FitH,     
						pdfpagemode=UseNone,      bookmarksopen=true]{hyperref}  
\usepackage[all]{hypcap}     

\newcommand{\comm}[1]{} 

\def\({\left(}
\def\){\right)}
\def\[{\left[}
\def\]{\right]}

\def\pd{{\partial}}

\def\One{{\hbox{ 1\kern-.8mm l}}}

\def\barray{\begin{array}}
\def\earray{\end{array}}
\def\be{\begin{equation}}
\def\ee{\end{equation}}
\def\bea{\begin{eqnarray}}
\def\eea{\end{eqnarray}}
\def\bal{\begin{align}}
\def\eal{\end{align}}

\numberwithin{equation}{section} 
\allowdisplaybreaks  

\makeatletter
\g@addto@macro\bfseries{\boldmath}
\makeatother

\definecolor{cardinal}{rgb}{0.6,0,0}
\definecolor{darkgreen}{rgb}{0,0.4,0}
\definecolor{golden}{rgb}{0.92, 0.7, 0}
\definecolor{midnight}{rgb}{0, 0, 0.5}
\definecolor{darkblue}{rgb}{0, 0, 0.7}
\definecolor{purple}{rgb}{0.5, 0, 0.5}

\def\Neql#1{{\cal N}\!=\!{#1}}

\def\IR{\mathbb{R}}

\def\IT{\mathbb{T}}

\def\ZZ{\mathbb{Z}}

\def\cB{{\cal B}}

\def\cD{{\cal D}}
\def\cF{{\cal F}}
\def\cG{{\cal G}}

\def\cL{{\cal L}}

\def\cN{{\cal N}}
\def\cP{{\cal P}}

\def\cO{{\cal O}}

\def\scp{{\sqrt{\cP}}}

\def\tF{{\tilde F}}
\def\tA{{\tilde A}}

\newcommand{\sig}[1]{{\sigma^{(#1)}}}

\usepackage{epsf}
\usepackage{cite}
\usepackage{fancyhdr}
\usepackage{bm}
\usepackage{enumerate} 

\usepackage{empheq} \empheqset{box=\fbox}
\usepackage[usenames,dvipsnames]{xcolor}

\numberwithin{equation}{section}  
\allowdisplaybreaks

\usepackage{graphicx}
\usepackage{tikz}
\usetikzlibrary{decorations.markings}
\tikzset{->-/.style={decoration={
			markings,
			mark=at position #1 with {\arrow{stealth}}},postaction={decorate}}}
\usepackage{adjustbox}
\usepackage{pgfplots}
\pgfplotsset{compat=1.11}
\usepgfplotslibrary{fillbetween}
\usetikzlibrary{intersections}
\usetikzlibrary{patterns}

\pgfdeclarelayer{bg}
\pgfsetlayers{bg,main}

\usetikzlibrary{calc}
\tikzset{
	samples=100,
}
\pgfplotsset{compat=1.11}

\pgfmathsetmacro\T{3.14}
\pgfmathsetmacro\A{0.2}
\pgfmathsetmacro\N{4}
\pgfmathsetmacro\D{\N*\T}

\begin{document}


\begin{flushright}

\end{flushright}

\vspace{3mm}

\begin{center}

{\huge {\bf Linearizing the BPS Equations with\\ \vspace*{5pt}Vector  and Tensor Multiplets}}

\vspace{14mm}

{\large
\textsc{ Nejc \v{C}eplak$^a$, Shaun Hampton$^a$, Nicholas P. Warner$^{abc}$}}
\vspace{12mm}

\textit{$^a$Universit\'e Paris Saclay, CNRS, CEA,\\
Institut de Physique Théorique,\\
91191, Gif-sur-Yvette, France
\\[12 pt]
\centerline{$^b$Department of Physics and Astronomy}
\centerline{and $^c$Department of Mathematics,}
\centerline{University of Southern California,} 
\centerline{Los Angeles, CA 90089, USA}}

\medskip

\vspace{4mm} 
%

{\footnotesize\upshape\ttfamily   nejc.ceplak, shaun.hampton  @ ipht.fr; warner @ usc.edu.} \\
\vspace{13mm}

\textsc{Abstract}

\end{center}

\begin{adjustwidth}{10mm}{10mm} 
 
\vspace{1mm}
\noindent

\noindent 
We analyse the BPS equations of $\cN = (1,0)$ supergravity theory in six dimensions coupled to  a vector and tensor multiplet.  
We show how these BPS equations can be reduced to a set of linear differential equations. 
This system  is triangular in that each layer of equations, while linear,  is quadratically sourced by the solutions of the previous layers. 
We examine several explicit examples and discuss the construction of new families of microstate geometries.
We expect that the result presented here will open up new branches of  superstrata  in which the momentum is encoded in a new class of charge carriers.

\end{adjustwidth}

\thispagestyle{empty}
\clearpage



\baselineskip=14.5pt
\parskip=3pt

\tableofcontents

\baselineskip=15pt
\parskip=3pt


\section{Introduction}
\label{sec:Intro}


Supersymmetric solutions play a very important role in supergravity, with applications ranging from holographic field theory to black-hole physics.  Such solutions are not only more stable but are also much easier to construct.  This is because  supersymmetry implies one must solve a {\it first-order system} of BPS equations rather than the full, second-order  equations of motion.  However, the BPS equations can still be extremely challenging because they are also, typically, non-linear.  

It was thus a remarkable surprise when it was shown that the BPS equations of five-dimensional $\Neql{2}$ supergravity coupled to vector multiplets \cite{Gauntlett:2003fk,Gutowski:2004yv,Bena:2004de} could be reduced to a triangular system of linear equations \cite{Bena:2004de,Bena:2005va, Bena:2007kg}. The first layer of this system consists of homogeneous, linear self-duality equations for the magnetic fields; the next layer involves linear equations for the electric potential with quadratic sources from the magnetic fields, and the final layer of the linear equations determines the angular-momentum vector of the metric in terms of the electric and magnetic fields.  Triangularity connotes the fact that, at each layer, the equations are linear but with (known) sources that are determined in the earlier layers of the system.  Indeed, such triangular structures were further developed and understood in terms of nilpotent orbits \cite{Bossard:2011kz}.

On a practical level, the breakthrough was that the non-linearities inherent in the BPS system could be re-organized into an algorithmically solvable linear system of equations whose only non-linearities appear in  source terms.%
\footnote{More precisely, the BPS equations always involve a ``zeroth-order'' layer that determines a spatial base geometry, and this ``zeroth-order'' layer  can be intrinsically non-linear.  In almost all known  BPS solutions, this base geometry is taken to be flat, or a Gibbons-Hawking geometry, but there are more general possibilities that involve  solving the ``zeroth-order''  non-linear equations.  The linearity of the BPS system that we discuss here is  all about solving the BPS excitations in an appropriately chosen  base geometry. We will discuss this further in Section \ref{sec:6DGravity}.}    This observation led to extensive and rapid development of the microstate geometry program  \cite{Bena:2005va, Bena:2006is, Bena:2006kb,Bena:2007kg,Bena:2007ju,Bena:2007qc, Bena:2008wt,Bena:2008nh,Bena:2008dw,Bena:2010gg} as a way to access and understand black-hole microstructure.

It was originally believed that the triangular, linear structure was a  special feature of a particular class of five-dimensional supergravities.   However, some years later it was shown that the BPS equations of six-dimensional $(1,0)$-supergravity coupled to tensor multiplets \cite{Gutowski:2003rg,Cariglia:2004kk}  could also be linearized \cite{Bena:2011dd,Giusto:2013rxa}.  Once again, this led to a breakthrough in the study of microstate geometries, enabling the explicit construction of  vast families of such geometries, including {\it superstrata}, which are the gravitational duals  are of momentum-carrying states in the D1-D5 CFT \cite{Giusto:2011fy, Lunin:2012gp, Giusto:2012yz,Bena:2015bea, Bena:2016agb, Bena:2016ypk,Bena:2017geu,Bena:2017xbt,Bakhshaei:2018vux,Ceplak:2018pws,Heidmann:2019zws,Heidmann:2019xrd,Shigemori:2020yuo,Mayerson:2020tcl,Houppe:2020oqp,Ganchev:2021iwy}. 

In this paper, we will analyze the BPS equations of six-dimensional $(1,0)$-supergravity coupled to a tensor multiplet and a vector multiplet, and show that, once again, the BPS equations have a triangular linear structure.     As one might expect, this work was also driven by developments in the microstate geometry program. 

The core precept of microstate geometries, and fuzzballs \cite{Bena:2022ldq}, is that horizons and singularities only arise in theories that do not possess sufficiently many degrees of freedom to describe black-hole microstructure.  It is therefore believed that, in string theory, gravitational collapse to the horizon scale will lead to a new state of matter that results in a fuzzball.  Since fuzzballs have no horizons, their  states can be measured and accessed by distant observers and so such objects do not suffer from an Information Paradox.  Microstate geometries are simply  the smooth, horizonless supergravity realizations of coherent fuzzball states.

The imperative for microstate geometries is that they are smooth, horizonless solutions, and that the space of such solutions should only degenerate into horizonless fuzzballs.  It was therefore a cause of concern when it became evident that a corner of the moduli space of superstrata appeared to result in a black hole with a macroscopic horizon.

Microstate geometries, and superstrata in particular, had met such challenges before: there appeared to be instabilities or limits in which they developed singularities or horizons.  However, time and time again, these pathological limits could be resolved within the fuzzball paradigm. The problems all arose as the result of freezing out crucial stringy degrees of freedom in the gravitational analysis, and, once the appropriate degrees of freedom are included, the dangerous limit, or instability, is resolved and new branches of  fuzzballs, or the moduli space of microstate geometries, are opened up.  

  As was shown in \cite{Bena:2022sge}, this is precisely what happens with the corner of the moduli space of superstrata that appears to result in a macroscopic black hole. First, the black-hole limit emerges from an inconsistent limiting process in which one eliminates momentum carrying modes while retaining the RMS value of their effects.   Moreover, the pathological limit can be resolved into a fuzzball, or {\it ``degenerate microstate solution,''} by incorporating new classes of momentum carrying excitations into the formulation of superstrata.
  
  From the perspective of supergravity, coherent combinations of these new momentum carriers source vector multiplets in the six-dimensional $(1,0)$ supergravity, and these more general classes of superstrata require at least six-dimensional $(1,0)$ supergravity coupled to both tensor {\it and vector} multiplets.   Hence the new imperative for this work:  the analysis of the BPS equations of this extended class of six-dimensional supergravities. 

 
In Section  \ref{sec:6DGravity}, we describe the relevant six-dimensional supergravity theory and its BPS equations.  In Section \ref{sec:Lin} we make field re-definitions that enable us to linearize the BPS equations, and we unpack their layered structure in which the linear equations in a given layer are sourced, non-linearly, by the solutions to the linear equations in preceding layers.   We present several explicit examples in Section~\ref{sec:Examples}, including the  three-charge solution with zero horizon area constructed in \cite{Bena:2022sge}.
Section \ref{sec:Summary} contains an ``Executive Summary'' of our results: The complete linearized Ansatz, the BPS equations, and the order in which they need to be solved. 
We make some final comments about possible further research  in Section~\ref{sec:Discussion}.
In Appendix~\ref{app:NoShift} we present a different form of the supersymmetric Ansatz and BPS equations that emphasises quantities that are invariant under the U(1) gauge symmetry.

\section{A review of six-dimensional  (1,0) supergravity}
\label{sec:6DGravity}

\subsection{Determining the relevant supergravity theory}
\label{ss:6Dtheory}

Minimal $(1,0)$ supergravity theory in six dimensions \cite{Nishino:1984gk, Nishino:1986dc,DAuria:1997caz} has eight supersymmetries, and the graviton multiplet contains the graviton, $g_{MN}$, two left-handed gravitini, and a self-dual tensor gauge field, $G^+_{MNP}$.    This theory can be thought of as the minimal result of reducing IIB supergravity on a $\IT^4$.  The tensor gauge field is then sourced by a D1 in six dimensions and a D5 wrapped on the $\IT^4$, giving a second ``effective D1'' in six dimensions.   Self-duality means that the six-dimensional dynamics of the D1 and D5 have been locked together.  

Minimal supergravity theory can be coupled to an arbitrary number of vector and tensor multiplets.  In a tensor multiplet, the  $3$-form  field strength  is necessarily anti-self-dual,%
\footnote{Self-duality or anti-self-duality correlates with the helicity of the supersymmetries, or gravitini, and the multiplet structure.  In our conventions, the supersymmetries are left-handed, self-dual tensors lie in the graviton multiplet, and so there can only be one such tensor field in the $(1,0)$ theory.  The additional tensor multiplets are then anti-self-dual.}
and there are two right-handed fermions, as well as one scalar.   A vector multiplet contains the vector, two left-handed fermions, and no scalar fields.

To describe the simplest independent brane excitations in the $\IT^4$ compactification of the D1-D5 system (or the S-dual NS5-F1 system) one must add a single (anti-self-dual) tensor multiplet.  The tensor gauge fields, $G^+_{MNP}$ and $G^-_{MNP}$, can be combined into a single, unconstrained tensor gauge field, $G_{MNP}$.  The electric components of  $G_{MNP}$ describe the D1 brane dynamics and the (independent) magnetic components describe the compactified D5 branes. This theory also has only a single scalar ``dilaton.''  

In the standard superstratum in the D1-D5 frame, the momentum carriers are  a combined set of NS5-F1 excitations that source the Kalb-Ramond field in IIB supergravity, and hence source another anti-self-dual tensor multiplet in six dimensions.%
\footnote{The anti-self-duality locks the F1 dynamics to that of the NS5 dynamics in a manner that preserves the supersymmetry of the D1-D5 system.}  (In the S-dual frame NS5-F1 frame, the momentum carriers are combined D5-D1 excitations.) In six dimensions, the resulting supergravity theory has a self-dual tensor gauge field, and two anti-self-dual tensor gauge fields and these transform as a vector of $SO(1,2)$.  There are two scalars, a dilaton and an axion, and these can be parametrized as the coset $SO(1,2)/SO(2)$.  One can, of course, couple more such multiplets, but the two-charge system with one set of  the ``original superstratum momentum carriers''  only requires two such anti-self-dual tensor multiplets. 

In the degenerate microstate solutions of  \cite{Bena:2022sge}, the  momentum carriers are  D0-D4 excitations on an NS5-F1  world-volume. In IIA supergravity, these D0-D4 excitations source the Kaluza-Klein vector field obtained from the $S^1$ compactification of M-theory.  Based on the observations in \cite{Bena:2022sge}, we believe that 
these excitations can be used to carry the momentum in new families of superstrata, and, to describe these, one will need to work  in the IIA frame (or in M-theory) and start from  the two-charge NS5-F1 system.
In six dimensions, the  essential ingredients to describe this system are contained in  $(1,0)$ supergravity coupled to a single (anti-self-dual) tensor multiplet to give the independent dynamics of the two-charge, NS5-F1, system plus a vector multiplet to describe the D0-D4  momentum carriers. 

One can, of course, couple more  vector multiplets and tensor multiplets.  Indeed, one can also  couple  non-Abelian vector multiplets to $(1,0)$ supergravity.  We expect that families of new momentum carriers will typically source Abelian vector fields, but there could be very interesting dynamics in which such fields interact with one another and this may occasion the need for the  more general non-Abelian vector multiplets.   However,  our goal here is to study the simplest BPS system associated with the new momentum carriers of  \cite{Bena:2022sge}.  We therefore restrict our focus to $(1,0)$ supergravity theory coupled to a single anti-self-dual tensor multiplet and a single vector multiplet.  Based on past experience, we anticipate that adding more tensor multiplets and  (Abelian) vector multiplets will only complicate our analysis by adding indices to fields but will not present any fundamentally new features to the BPS analysis that we will perform here.

\subsection{The fields and action}
\label{ss:Fields}

Our analysis of the BPS equations will start from  the analysis in \cite{Gutowski:2003rg, Cariglia:2004kk}, and so we will adopt most of their conventions.%
\footnote{However, in contrast to  \cite{Gutowski:2003rg, Cariglia:2004kk} (and \cite{Bena:2011dd}), we use a mostly positive metric signature, similar to \cite{Salam:1984cj}. Where possible, we follow the conventions of \cite{Giusto:2013rxa}. Specifically, we have rescaled the fields as $\phi^{\rm here}= \sqrt2\, \phi^{\rm there}$, $A^{\rm here} = \sqrt 2\, A^{\rm there}$ , and $B^{\rm here}= 2B^{\rm there}$,
which, through \eqref{eq:FieldStrengthsDef},  implies that $   F^{\rm here}= \sqrt 2\, F^{\rm there}$ and $  G^{\rm here}= 2\, G^{\rm there} $. The gravitational coupling is set to  $\kappa = 1$ throughout.}
The bosonic sector of minimal supergravity coupled to a vector multiplet and a single anti-self-dual tensor multiplet consists of the   graviton, $g_{MN}$; an unconstrained tensor gauge field, $G_{MNP}$, with a potential, $B_{MN}$; a Maxwell field, $F_{MN}$, with vector potential, $A_M$; and a scalar dilaton, $\phi$.
In particular, we have:
\begin{align}
	\label{eq:FieldStrengthsDef}
	F ~\equiv~ d A\,,\qquad G ~\equiv~ dB + F \wedge A\,.
\end{align}

The bosonic action has the Lagrangian density
\begin{align}
	\label{eq:Lagrangian}
	e^{-1} \, \cL ~=~ \frac14\,R -\frac14\, \nabla_M\phi\, \nabla^M\phi- \frac18\, e^{2 \phi}\, G^2-\frac14\, e^\phi \, F^2\,,
\end{align}
where we define the square of a $p$-form, $X_p$,  by:
\begin{align}
	\label{eq:FormContr}
	X_p^2 ~\equiv~~ \frac{1}{p!}\, X_{M_1, M_2, \ldots M_p}\,X^{M_1, M_2, \ldots M_p}\,.
\end{align}
The  bosonic equations of motion and Bianchi identities  are given by:
\begin{subequations}
	\label{eq:BosEOM}
	\begin{align}
		&R_{MN} = \nabla_M\phi \, \nabla_N \phi + \frac14 \,e^{2\phi} \,\left( G_{M\alpha \beta}\,G_N{}^{\alpha \beta} - g_{MN}\, G^2\right)+ \frac14 \, e^\phi\, \left(4 F_{M\alpha}\,F_N{}^{\alpha}- g_{MN}\, F^2\right),\\
		&\nabla^2 \phi ~=~ \frac12 e^{2\phi}\,G^2 + \frac12 e^\phi \, F^2\,,\\
		&\label{eq:BosEOMG}
		d\left(e^{2\phi}*_6 G\right)  ~=~ 0 \,,\qquad \qquad\qquad\qquad \quad
		dG  ~=~ F\wedge F\,,\\
		&d\left(e^\phi *_6 F\right) \,\,~= ~e^{2\phi}*_6G \wedge F\,, \qquad \qquad 
		\, dF ~=~ 0\,.
	\end{align}
\end{subequations}
where the conventions for the Hodge dual are those of \eqref{eq:HDDef}.

\subsection{The BPS equations}
\label{ss:BPS}

The BPS equations for the minimal supergravity coupled to one tensor multiplet were first analysed in  \cite{Gutowski:2003rg}, and this analysis was extended to include vector multiplets in \cite{Cariglia:2004kk}.   The underlying linear structure for the minimal theory coupled to tensor multiplets was only uncovered in \cite{Bena:2011dd,Giusto:2013rxa}, and our goal  is to reveal the underlying linear structure of the BPS equations given in \cite{Cariglia:2004kk}.

The BPS analysis of \cite{Gutowski:2003rg}  fixed the form of the metric, and this is unmodified by the inclusion of vector and tensor multiplets \cite{Cariglia:2004kk,Bena:2011dd,Giusto:2013rxa}.  Indeed, the metric can be written as
\begin{align}
	\label{eq:MetAns}
	ds_6^2  &~=~-\frac{2}{\sqrt{\cP}}\,(d v+\beta)\Big[d u+\omega + \frac{\widetilde\cF}{2}(d v+\beta)\Big]~+~\sqrt{\cP}\,d s^2_4(\cB)\,,
\end{align}
where
\begin{align}
	d s^2_4(\cB) &~= ~h_{mn}\, dx^m\, dx^n\,,
	\label{basemet}
\end{align}
is the metric on a four-dimensional base space, $\cB$. 
The fields, $\omega$ and $\beta$, are one-forms on this base space, and $\widetilde \cF$ and $\cP$ are scalars. 
The supersymmetry implies the existence of a  null Killing vector field, which we have taken to be $\frac{\partial}{\partial u}$.  Generically, the metric and all the fields
can be functions of $v$ and the base-space coordinates, $x^m$, and are only  independent of the coordinate $u$.

The coordinates $u$ and $v$ are the canonical null coordinates, and it is useful to introduce space and time coordinates, $(t,y)$, defined by:
\begin{equation}
u ~\equiv~\frac{1}{\sqrt{2}} \, ( t ~-~y  ) \,, \qquad  v ~\equiv~\frac{1}{\sqrt{2}} \, (t ~+~y )\,, 
\label{uvtyreln}
\end{equation}
where $y$ is periodically identified on a circle, $S^1(y)$,  with:
\begin{equation}
\label{yperiod}
y ~\equiv~ y ~+~ 2 \pi \,R_y\,.
\end{equation}
It is convenient to introduce a null frame for the metric \eqref{eq:MetAns}
\begin{align}
	\label{eq:NullFrame}
	ds_6^2 ~=~ -2\,e^+ \, e^- ~+~ \delta_{ab}\,e^a \, e^b\,,
\end{align}
where
\begin{align}
	\label{eq:6DVielbeine}
	e^+ \equiv \frac{1}{\sqrt{\cP}}(dv+ \beta) \,, \qquad e^- \equiv du + \omega + \frac{\widetilde\cF}{2} \sqrt{\cP} e^+\,, \qquad e^a \equiv \cP^{\frac14} \, \tilde e^a\,,
\end{align}
and $\tilde e^a\equiv\tilde e^a{}_m\, dx^m$, $a = 1,\dots,4$, define a vierbein for the base space.

The  conditions for supersymmetry imply that the base space $\cB$ is ``almost hyper-K\"ahler.'' That is, there are a set of three anti-self dual two-forms, $J^A$, with $A= 1,2,3$, on the base space that satisfy a quaternionic algebra%
\footnote{Note that compared to \cite{Cariglia:2004kk}, we flip the sign of the anti self-dual two forms $J^A_{\rm here} = - J^A_{\rm there}$, just as in \cite{Bena:2011dd}.}
\begin{subequations}
	\label{eq:AlmHypKahAlg}
	\begin{align}
		*_4 J^A ~&=~ - J^A\,,\\
		(J^A)^m{}_{n}(J^B)^n{}_{p}~&=~ \epsilon^{ABC}(J^C)^m{}_{p}-\delta^{AB}\delta^m_p\,,\\
		J^A \wedge J^B ~&= ~- 2\delta^{AB}\, {\rm vol}_4\,,
	\end{align}
\end{subequations}
and obey the differential relation 
\begin{align}
	d_4 J^A &~=~ \pd_v\left(\beta \wedge J^A\right)\,,
\end{align}
where  $d_4$ denotes the exterior derivative restricted to the base space, $\cB$. 

We also use the non-standard conventions of \cite{Gutowski:2003rg,Cariglia:2004kk} for the Hodge dual, $*_D$,  in $D$-dimensions.  That is,  for a $p$-form, $X_p$, we define:
\begin{align}
	\label{eq:HDDef}
	*_D X_p  ~\equiv~ \frac{1}{p! (D-p)!}\, \epsilon_{m_1\ldots m_{D-p}, n_{D-p+1} \ldots n_D}\, X^{n_{D-p+1} \ldots n_D}\, e^{m_1}\wedge \ldots e^{m_{D-p}}\,,
\end{align}
and throughout  we choose the orientation
\begin{align}
	\epsilon^{+-1234}  ~=~\epsilon^{1234} ~=~ 1\,,
\end{align}
using the  frame  in \eqref{eq:6DVielbeine}.

Supersymmetry  imposes a self-duality condition on the one-form $\beta$:
\begin{align}
	\label{eq:BetaEq}
	\cD \beta & ~=~ *_4\cD \beta\,,
\end{align}
where $\cD$ is  a differential operator that  acts on an arbitrary form, $X$,  via:
\begin{align}
	\label{eq:DefcalD}
	\cD X ~\equiv~ d_4 X ~-~ \beta \wedge\dot{X}\,,
\end{align}
with $\dot X \equiv \pd_v X$.%
\footnote{We will make use of the following identities involving  \eqref{eq:DefCalG}
	\begin{align*}
		&d X_p  = (dv + \beta) \wedge \dot X_p+ \cD X_p\,,& & 
		\cD X_p = d_4 X_p - \beta \wedge \dot X_p\,,\\
		& \cD^2 X_p = - \cD \beta \wedge \dot X_p\,,&& 
		\pd_v\left(\cD X_P\right)  = \cD \dot X_p - \dot \beta \wedge \dot X_p\,,
	\end{align*}
	for an arbitrary $p$-form $X_p$.
	Furthermore, one can show that the differential operator behaves as an exterior derivative:  \mbox{$\cD(X_p\wedge Y) = \cD X_p \wedge Y + (-1)^p X_p\wedge \cD Y$}.}
This operator is covariant under diffeomorphisms that  preserve the form of the metric \eqref{eq:MetAns} \cite{Giusto:2013rxa,Bena:2015bea}:
\begin{align}
	v \to v + V(x^m)\,, \qquad \beta \to \beta - d_4 V\,.
\end{align}

The  gauge fields are also greatly constrained by supersymmetry. 
As noted in  \cite{Cariglia:2004kk, Bena:2011dd}, the three-form gauge field strength and its Hodge dual can be decomposed as:
\begin{eqnarray}  
	e^{\phi} \, G  &=&\, *_4 \big( \cD \scp  + \scp \, \dot \beta  -  \, \scp\, \cD \phi \big) \nonumber  \\
	&&  \,-\,  e^+ \wedge e^-  \wedge \big( \frac{1}{\scp}\cD\scp  +  \, \dot \beta  + \,  \cD \phi \big) \nonumber  \\
	&& \,+\,  e^+ \wedge \big(  \cP \psi  -   (\cD\omega)^-  +K    \big)   +    \frac{1}{\scp}  \, e^- \wedge \cD\beta \,,  \label{Gform}   
\end{eqnarray}
and
\begin{eqnarray}  
	e^{ \phi} \,*_6 G  &=& \,- *_4 \big( \cD\scp + \scp \, \dot \beta  +   \, \scp\, \cD \phi \big)  \nonumber  \\
	&&  +\,  e^+ \wedge e^-  \wedge \big( \frac{1}{\scp} \cD\scp  +  \, \dot \beta  -  \,  \cD \phi \big)  \nonumber  \\
	&&  +\,e^+ \wedge \big(- \cP \psi  + \,  (\cD\omega)^- +  K    \big)   - \frac{1}{\scp}  \, e^- \wedge \cD\beta \,. \label{dualGform}   
\end{eqnarray}
where
\begin{align}
	(\cD \omega)^\pm ~\equiv~  \frac{1}{2}\, \left(\cD \omega \pm *_4 \cD \omega\right) \label{ompmdefn}   \,,
\end{align}
$K$ is a self-dual two form on the base space:
\begin{align}
	K ~=~ *_4 K\,,
\end{align}
and 
\begin{align}
	\psi ~\equiv~  \frac{1}{8}\, \epsilon_{ABC}\, (J^A)^{mn}\, (\dot J^B)_{mn}\, J^C\,,
\end{align}
which is anti-self dual due to the properties, \eqref{eq:AlmHypKahAlg}, of the $J^A$ . 
It is also convenient to define a self-dual two-form:%
\footnote{To go to the field definitions of \cite{Cariglia:2004kk}, one needs to take $K^{\rm here} = 2 K^{\rm there}$, $\cG^{\rm here} = 2\, \scp\, \cG^{\rm there}$. Similarly, due to a rescaling of the  two-form gauge field strength, its components are rescaled as  $\omega_F^{\rm here} = \frac{\sqrt2}{\scp}\,\omega_F^{\rm there}$ and $\tF^{\rm here} = {\sqrt2}\,\tF^{\rm there}$.}
\begin{align}
	\label{eq:DefCalG}
	\cG ~=~ \left(\cD\omega\right)^+ + \frac{\widetilde\cF}{2}\,\cD\beta\,.
\end{align}

The vector fields  are also constrained in that two-form field strength must have the form  \cite{Cariglia:2004kk}:
\begin{align}
	\label{eq:Fform}
	F ~=~ e^+ \wedge \scp\, \omega_F ~+~ \tF\,,
\end{align}
where $\omega_F$ is a one-form on the base space and $\tF$ is a self-dual two form on the four-dimensional base:
\begin{align}
	\label{eq:SDtF}
	*_4 \tF ~=~ \tF\,.
\end{align}

Inserting the decomposition of the metric  and gauge fields into the bosonic equations of motion \eqref{eq:BosEOM} reduces to BPS equations for the  quantities appearing in the components of the fields.
In turn, this means that solving this system of equations  guarantees that the bosonic equations of motion are satisfied  \cite{Cariglia:2004kk}. 

The closure of the two-form field strength leads to the constraint
\begin{align}
	\label{eq:OmegaFeq0}
	\cD\left(\cP *_4 \omega_F\right) ~=~ - 2  \cD\phi \wedge \left(\cP *_4 \omega_F\right) ~+~  \tF \wedge \left(K-  \cG\right)\,,
\end{align}
and inserting the Ansatz into the equations involving the three-form field strengths gives
\begin{subequations}
	\label{eq:DiffCons1}
	\begin{align}
		\label{eq:Bianchi1}
		&\cD\left(\frac{e^{ \phi}}{\scp}\left(K- \cG - \cP \psi\right) \right)+  \pd_v *_4\left(\cD \left(e^{\phi}\, \scp\right) + e^{ \phi}\, \scp\,\dot \beta\right)\nonumber\\
		& - \frac{e^{\phi}}{\scp}\dot \beta \wedge \left(K- \cG - \cP \psi\right)= 0\,,\\
		\label{eq:Bianchi2}
		&-\cD\left(\frac{e^{- \phi}}{\scp}\left(K+ \cG - \cP \psi\right) \right)+  \pd_v *_4\left(\cD \left(e^{- \phi}\, \scp\right) + e^{- \phi}\, \scp\,\dot \beta\right)\nonumber\\
		& + \frac{e^{- \phi}}{\scp}\dot \beta \wedge \left(K+ \cG - \cP \psi\right)= 2 \, \omega_F \wedge \tF \,,\\
		\label{eq:BEOM1}
		& \,\cD *_4 \left(\cD \left(e^{ \phi}\, \scp\right) + e^{ \phi}\, \scp\,\dot \beta\right) - \frac{ e^{ \phi}}{\scp}\left(K -  \cG\right) \wedge \cD \beta=0\,,\\
		\label{eq:BEOM2}
		& \cD *_4 \left(\cD \left(e^{- \phi}\, \scp\right) + e^{- \phi}\, \scp\,\dot \beta\right) + \frac{e^{- \phi}}{\scp}\left(K + \cG\right) \wedge \cD \beta=  \tF \wedge \tF\,.
	\end{align}
\end{subequations}

As is typical of BPS equations, a solution  of the foregoing first-order system automatically  solves all but one component of  the equations of motion.    The remaining equation appears in the  $vv$-component of the  Einstein equation, and this leads  to the only second order equation that one needs to solve in order to find the full classical solution:
\begin{align}
	\label{eq:EFE++}
	&*_4 \cD\left(*_4 \left[ \dot \omega + \frac{\widetilde \cF}{2}\dot \beta - \frac{1}{2} \cD \widetilde \cF\right]\right)\nonumber \\*
	& = \frac{\cP}{2} h^{mn} \pd_v^2 \left(\scp h_{mn}\right) + \frac14 \pd_v \left(\scp  h^{mn}\right) \pd_v \left(\scp  h_{mn}\right) - 2 \dot \beta_m \left[ \dot \omega + \frac{\widetilde \cF}{2}\dot \beta - \frac{1}{2} \cD \widetilde \cF\right]^m\nonumber\\*
	&  - \frac{1}{2 \cP} \left(\cD\omega + \frac{\widetilde \cF}{2}\, \cD\beta\right)^2 + \frac{1}{8\cP}\left(K + \cP \psi + \left(\cD \omega\right)^-\right)^2 +  \cP \,\dot\phi^2+ \scp \,e^{ \phi}\, \omega_F^2\,,
\end{align}
where the squares of forms on the base, $\cB$, are defined by \eqref{eq:FormContr}  but  with the indices raised by $h^{mn}$, the  inverse of the base-space metric.

\section{Linearization}
\label{sec:Lin}

While the BPS equations presented above may be simpler than the full bosonic equations of motion, they are still highly non-trivial. 
However, following the methodology of \cite{Bena:2011dd,Giusto:2013rxa},  we now  show that these BPS equations   can be largely linearized in terms of a slightly more complicated set of layers. 

As in earlier linearizations of BPS systems   \cite{Bena:2004de,Bena:2005va, Bena:2007kg,Bena:2011dd,Giusto:2013rxa}, there is always a  ``zeroth layer'' of BPS equations that sets the background, or base  geometry upon which the linear equations are to be solved.   For six-dimensional supergravity, this zeroth layer determines the geometry of the base: its metric, (\ref{basemet}), with its ``almost hyper-K\"ahler'' structure \eqref{eq:AlmHypKahAlg}, and the geometry of the fibration, (\ref{eq:BetaEq}).  The general solutions of this non-linear system remain unknown  and so one makes simplifying assumptions.  Indeed,  one typically assumes that this background geometry is $v$-independent and then the metric (\ref{basemet}) is  simply hyper-K\"ahler,%
\footnote{One should, of course, remember that this hyper-K\"ahler metric does not need to be Riemannian:  it is allowed to be ambi-polar  \cite{Giusto:2004kj,Bena:2005va,Berglund:2005vb, Bena:2007kg}.} and the equation for the fibration vector, $\beta$, becomes a linear self-duality equation.  We also note that  taking the base metric to be flat $\IR^4$ still leads to a wealth of superstratum solutions.

However, we will not make any of these assumptions here.  We start by merely assuming that one has fixed the base metric  (\ref{basemet})  and fibration vector, $\beta$, in accordance with  \eqref{eq:AlmHypKahAlg}  and (\ref{eq:BetaEq}). We now show that the remaining BPS equations can be linearized.

\subsection{Ansatz for the linearization}

The algorithm for linearization involves a careful separation of the  electric potentials from the magnetic degrees of freedom.  The BPS equations generally mean that the electric potentials then appear in the metric warp factors and in the scalar fields. 

With this in mind, we define the following scalar fields as a combination of the dilaton and the scalar warp factor:
\begin{align}
	Z_1 ~&\equiv~ \scp\, e^{-\phi}\,,&
	Z_2 ~&\equiv~\scp\, e^{\phi}\,.  \label{ZDefns}
\end{align}
or conversely
\begin{align}
	\cP &~=~ Z_1 \, Z_2\,, & e^{2\,\phi}& ~=~ \frac{Z_2}{Z_1} \quad \Leftrightarrow \quad \phi ~=~ \frac{1}{2} \log\left(\frac{Z_2}{Z_1}\right)\,.
\end{align}

The new ingredient in the theory, compared to \cite{Bena:2011dd, Giusto:2013rxa}, is the additional U(1) gauge field coming from the vector multiplet. 
The self-duality condition \eqref{eq:SDtF} is manifestly linear, but the other  BPS equation \eqref{eq:OmegaFeq0}, while seemingly linear in $\tF$ and $\omega_F$, still requires us to show that the other pieces of  this equation can be obtained independently of $\tF$ and $\omega_F$.

Following the standard process, we decompose  the gauge-field potential, $A$, into electric and magnetic parts:
\begin{align}
	\label{eq:AAns}
	A ~=~ \frac{Z_A}{Z_2}\, \left(dv + \beta\right) - \tA\,,
\end{align}
where   $Z_A$ is a conveniently chosen scalar potential,  and $\tA$ is a one-form on the four-dimensional base.
As is evident from	(\ref{eq:Fform}) and (\ref{eq:6DVielbeine}), supersymmetry means that $A$ has no $du$ component \cite{Cariglia:2004kk}. 

Note that in writing \eqref{eq:AAns}, we have undone a choice of gauge that was made in \cite{Cariglia:2004kk}, in which $Z_A$ was set to zero.
Using $F = dA$ we can then read off
\begin{align}
	\label{eq:OmegaTildeF}
	\omega_F ~=~-\dot \tA + \frac{Z_A}{Z_2} \, \dot \beta - \cD \left(\frac{Z_A}{Z_2}\right)\,,\qquad 
	\tF~ =~- \cD \tA +\frac{Z_A}{Z_2}\, \, \cD \beta\,.
\end{align}
We see that since $\cD\beta$ is self dual \eqref{eq:BetaEq}, the self-duality condition of $\tF$ implies the self-duality of $\cD \tA$.

Next, we define the following two-forms which appear in the BPS equations determining the components of the three-form field strengths
\begin{subequations}
	\label{eq:ThetaDefns} 
	\begin{gather}
		\Theta^1 ~\equiv~ \frac{e^{\phi}}{\scp} (-K+ \cG + \cP \psi)\,, \\*
		\Theta^2+ \frac{Z_A^2}{Z_2^2}\, \cD\beta -2\, \frac{Z_A}{Z_2}\, \cD \tA ~\equiv~  \frac{e^{-\phi}}{\scp}\,  \left(K+\cG + \cP \psi  \right)\,,
	\end{gather}
\end{subequations}
where,  in comparison to  \cite{Bena:2011dd, Giusto:2013rxa}, the definition of $\Theta^2$ includes additional terms involving the  U(1) gauge field. 
This modification can be traced  to the definition of the three-form gauge field \eqref{eq:FieldStrengthsDef}, which is no longer  closed because of the  U(1) gauge field. (See  (\ref{eq:BosEOM}).)
While the new terms in $\Theta^2$ may seem rather convoluted, we will see that they are responsible for substantial cancellations in the BPS equations. 
We will also make some comments about the physical importance of these terms in Section~\ref{ssec:Shift}.

Inverting the definitions \eqref{eq:ThetaDefns} gives:
\begin{subequations}
	\label{eq:DefnsInv}
	\begin{align}
		\cG ~& =~ \frac12\,Z_1\,\Theta^1+ \frac{1}2\,Z_2 \,\left(\Theta^2+ \frac{Z_A^2}{Z_2^2}\, \cD\beta -2\, \frac{Z_A}{Z_2}\, \cD \tA\right)- Z_1\,Z_2\,\psi \,,\\*
		K ~&=~ -\frac12\, Z_1\,\Theta^1+ \frac12\,Z_2 \left(\Theta^2+ \frac{Z_A^2}{Z_2^2}\, \cD\beta -2\, \frac{Z_A}{Z_2}\, \cD \tA\right)\,. 
	\end{align}
\end{subequations}
Note that $\cG$, $K$, $\cD\beta$, and $\cD \tA$ are all self dual two-forms on the base space, which implies that the $\Theta^j$ are almost self-dual:
\begin{eqnarray}
	*_4 \Theta^1   &=&  \Theta^1  -  2 \,  Z_2\,  \psi  \,,  \label{eq:Theta1dual} \\
	*_4 \Theta^2   &=&  \Theta^2  -  2 \,  Z_1\,  \psi  \,, \label{eq:Theta2dual}
\end{eqnarray}
where the anti-self-dual parts of the $\Theta^j$ are proportional to $ \psi$.

\subsection{Deriving the linear BPS equations}

The primary purpose of all  these  field re-definitions is to simplify the BPS equations. 
To that end, we simply insert the decompositions into the original equations derived in \cite{Cariglia:2004kk} and uncover a hidden linear and layered structure.

We begin by analysing \eqref{eq:Bianchi1} and \eqref{eq:BEOM1}, which do not involve any contributions from the U(1) field.
In terms of the new variables, the equations reduce exactly as in \cite{Bena:2011dd}:
\begin{subequations}
	\label{eq:Sub10}
	\begin{gather}
		\label{eq:Z2eq1}
		\cD *_4 \left[ \cD Z_2 + Z_2 \,\dot \beta\right] ~+~ \Theta^1 \wedge \cD \beta ~=~ 0\,,\\
		d_4 \Theta^1 ~=~ \pd_v \left[ \beta \wedge \Theta^1 +  *_4 \left(\cD Z_2 + Z_2\, \dot \beta\right)\right]\,.
	\end{gather}
\end{subequations}
These equations only involve $Z_2$ and $\Theta^1$, which, when combined with the self-duality condition \eqref{eq:Theta1dual} suffice to completely determine these two quantities provided the base space metric and the one-form $\beta$ are known. This is the {\it first layer of the BPS system}.

The equations for the U(1) gauge field also greatly simplify.  Indeed (\ref{eq:OmegaFeq0}) reduces to 
\begin{align}
	\label{eq:OmegaFeq2}
	2\, \cD Z_2 \wedge *_4\omega_F ~+~ Z_2\, \cD *_4 \omega_F~=~ - \tF \wedge \Theta^1\,.
\end{align}
Since $Z_2$ and $\Theta^1$  are now known solutions of the first layer,  \eqref{eq:Sub10},  this equation, together with the equation   \eqref{eq:SDtF},   constitute a  linear  system for the vector field.

In practice, it might be more convenient to solve directly for the components of the gauge field potential. 
From the decomposition \eqref{eq:OmegaTildeF}, we see that the self-duality condition \eqref{eq:SDtF} 
translates into a condition on $\tA$ 
\begin{align}
	\label{eq:SDcDA}
	*_4 \cD \tA   ~=~ \cD \tA\,.
\end{align}
Similarly, \eqref{eq:OmegaFeq0} becomes 
\begin{align}
	\label{eq:EqZA}
	*_4 \cD Z_A \wedge \dot \beta ~+~2\, \cD Z_2 \wedge *_4 \dot \tA ~+~ Z_2\,  \cD * \dot\tA ~+~ \cD *_4 \cD Z_A~=\,  -\cD \tA \wedge \Theta^1\,.
\end{align}
While this equation is  not written in terms of gauge invariant quantities, it is on a similar footing as \eqref{eq:Z2eq1} and, most importantly involves the Laplacian on $Z_A$.  Indeed, if  $\beta$ and $\tA$ are $v$-independent then this equation is precisely the same as \eqref{eq:Z2eq1}.

We note that $A$ has five non-vanishing components \eqref{eq:AAns}, and a U(1) gauge invariance. The four physical components are determined  by solving the four equations  \eqref{eq:SDcDA} and \eqref{eq:EqZA}.  This is  the {\it second layer of the BPS system}.

We now turn to the remaining two equations of \eqref{eq:DiffCons1} whose right hand sides include quadratic terms in the U(1) gauge field. 
By inserting the field re-definitions into these equations, they become
\begin{subequations}
	\label{eq:Sub20}
	\begin{gather}
		\cD *_4 \left[ \cD Z_1 + Z_1 \,\dot \beta\right] ~+ ~
		\Theta^2 \wedge \cD \beta ~=~ \cD \tA \wedge \cD \tA\,,\\
		d_4 \Theta^2 ~=~ \pd_v \left[ \beta \wedge \Theta^2 +  *_4 \left(\cD Z_1 + Z_1\, \dot \beta\right)\right]~-~ 2 \dot \tA  \wedge \cD \tA \,,
	\end{gather}
\end{subequations}
which are equivalent to \eqref{eq:Sub10}, up to the  quadratic terms in the U(1) fields.   Once again note that the only unknowns here are $(Z_1, \Theta^2)$,  and that these equations are also linear. This is  the {\it third layer of the BPS system}.

This highlights a fundamental difference from the D1-D5 system \cite{Bena:2011dd,Giusto:2013rxa}.
There the ``first layer'' of the BPS equations determined both the pairs $(Z_1,\Theta^2)$ and $(Z_2, \Theta^1)$: these two sets of fields are entirely on the same footing.
Adding a vector multiplet breaks this equivalence and imposes a strict order in which one must proceed to retain  linear  BPS equations. 
Nonetheless, the general idea remains: Solving the BPS equations in the right order guarantees that one only needs to solve linear differential equations with the solution to the previous set of equations acting as, at most, quadratic sources. 

One is left with determining the one-form $\omega$ and the scalar $\widetilde \cF$, which contain the information about the angular momentum and the momentum carried by the geometry \cite{Myers:1986un}. 
It simplifies the result if we shift the scalar function, $\widetilde \cF$, and define:
\begin{align}
	\label{eq:cFShift}
	\widetilde \cF  ~\equiv~ \cF ~+~ \frac{Z_A^2}{Z_2}\,. 
\end{align}
This shift embodies the fact that the scalar function, $Z_A$, explicitly appears in the metric, 
and this re-definition decouples these $Z_A^2$ terms from the BPS equations. We will comment on the physical significance of this shift in Section~\ref{ssec:Shift}. 

To determine $\omega$, one can invert the defining equation for $\cG$, \eqref{eq:DefCalG}, to obtain 
\begin{align}
	\label{eq:OmegaSDEq}
	\cD \omega + *_4 \cD \omega ~=~  Z_1\, \Theta^1 + Z_2\, \Theta^2- 2\, Z_A\, \cD\tA - \cF \,\cD\beta - 2\, Z_1\,Z_2\,\psi \,,
\end{align} 
which shows that the contribution from the U(1) field is on the same footing as the contributions from the $Z_I$ and $\Theta^I$ fields. 

Finally, the residual  Einstein equation can be re-written as:
\begin{align}
	\label{eq:Lay2eq2New}
	&*_4 \cD *_4   L+ 2 \,\dot \beta^m \,  L_m \nonumber\\*
	&=- \frac12 *_4 \left( \Theta^1 - Z_2 \, \psi\right)\wedge \left(\Theta^2 - Z_1 \, \psi\right) +  \ddot Z_1 \, Z_2 + \dot Z_1\, \dot Z_2 + Z_1 \, \ddot Z_2 \nonumber \\*
	&\quad +\frac12 Z_1 \,Z_2 *_4  \psi\wedge  \psi +  *_4  \psi \wedge \cD \omega- \frac14 Z_1\, Z_2\, \dot h^{mn}\,\dot  h_{mn} + \frac12 \pd_v \left[Z_1 \, Z_2 \,h^{mn}\dot h_{mn} \right]\nonumber\\*
	& \quad + Z_2 \,\dot \tA^2 +  \dot\tA_m \left[\cD Z_A\right]^m\,,
\end{align}
where we have defined
\begin{align}
	\label{eq:LDef}
	L ~\equiv~  \dot \omega + \frac{\cF}{2} \,\dot \beta- \frac12\, \cD  \cF+ Z_A\, \dot\tA\,.
\end{align}
Note that the raising and lowering of indices is done with the base metric, $h_{mn}$, and, in particular, one can re-cast the last source term purely in terms of differential forms as  
\begin{align}
	\dot\tA_m \left[\cD Z_A\right]^m ~=~ - *_4\left(\dot \tA\wedge *_4 \cD Z_A\right)\,.
\end{align}
Observe that  given solutions to all the previous layers of the BPS system,  the  equations \eqref{eq:OmegaSDEq} and \eqref{eq:Lay2eq2New} are linear in $\omega$ and $\cF$. This defines  the {\it fourth, and final layer of the BPS system}.

It is interesting to observe that the vector field contributions in \eqref{eq:Lay2eq2New} are two-fold. 
There are two explicit terms appearing on the right hand side, both of which act as sources for $L$. 
But $L$ also contains the quadratic contribution $Z_A\,\dot \tA$. 
In that sense, one can clearly see a close parallel  between $\tfrac12\,\cF\, \dot\beta$ and $Z_A \, \dot \tA$ in \eqref{eq:LDef} which extends to $\tfrac12\, \cF\, \cD \beta$ and $Z_A\, \cD \tA$ in  \eqref{eq:OmegaSDEq}.  As we will discuss in Section \ref{ssec:Shift}, this  parallel is entirely natural from the perspective of a five-dimensional compactification of M-theory.

To summarize, the BPS equations can be organised into a ``triangular linear system.''  That is, they can be organised into layers of linear equations that can be solved in succession and in which the sources for a given layer are quadratic in the solutions to earlier layers.   
However,  the U(1) gauge field adds to the complexity: the original ``first layer'' in \cite{Bena:2011dd}  is now split into three sublayers which need to be solved in a particular order:
Once one has fixed the base space metric and the fibration, $\beta$, one first needs to solve \eqref{eq:Sub10}  and determine $Z_2$ and $\Theta^1$, followed by solving \eqref{eq:SDcDA} and \eqref{eq:EqZA} for the components of the U(1) gauge field. 
The latter fields then act as sources in the equations (\ref{eq:Sub10}) that determine $Z_1$ and $\Theta^2$. 
Finally, once all quantities connected to the gauge fields are determined, one needs to solve \eqref{eq:OmegaSDEq} and \eqref{eq:Lay2eq2New}, which determine $\omega$ and $\cF$ and thus the angular and linear momentum of the solution. 
%

\subsection{Summary: the Ansatz in linearized form}
\label{ssec:LinForm}

We now put all of the elements together to  give a simplified form of the BPS Ansatz. 

We first build the shift (\ref{eq:cFShift}) into the metric:
\begin{align}
	\label{eq:MetAnsNew}
	ds_6^2  ~=~-\frac{2}{\sqrt{Z_1\, Z_2}}\,(d v+\beta)\,\left[d u+\omega + \frac{1}{2}\,\left(\cF + \frac{Z_A^2}{Z_2}\right)(d v+\beta)\right]~+~\sqrt{Z_1\, Z_2}\,d s^2_4\,,
\end{align}

The signs of the $g_{vv}$ terms are very interesting.  For BPS solutions $\cF$ is typically negative.  This is evident if one compactifies to five dimensions, where $-\cF$ becomes an electric potential for the KK gauge field.  Thus $Z_A$ contributes negatively to such a charge.  It was the 
 interplay between these two contributions that allowed for  cancellations and can cause an inherently three-charge geometry to have a zero size horizon area \cite{Bena:2022sge}.

The  dilaton is given by a simple ratio of the scalar functions 
\begin{align}
	\label{eq:DilatonAnsatz}
	e^{2\phi} & ~=~ \frac{Z_2}{Z_1}\,,
\end{align}
while the one-form gauge potential is decomposed as 
\begin{align}
	A & ~=~ \frac{Z_A}{Z_2}\,(dv + \beta) - \tA\,,
\end{align}
with the associated gauge field strength $F = dA$ being given  by 
\begin{align}
	F ~ =~  (dv+ \beta) \wedge \omega_F + \tF\,,
\end{align}
where the explicit relation between the components of $A$ and $F$ are given in \eqref{eq:OmegaTildeF}.

The supersymmetric Ansatz three-form gauge field strengths \eqref{Gform} and its dual \eqref{dualGform} can be simplified to%
\footnote{The relative minus sign in \eqref{eq:HDG} compared to the analogous expression in \cite{Bena:2011dd} arises due to the difference in the signature of the metric.} 
\begin{subequations}
	\label{eq:GaugeNew}
	\begin{align}
		G ~&=~ d\left[ -\frac{1}{Z_2}\, (du +\omega)\wedge (dv + \beta)\right] ~+~ \widehat G_2\,,\\
		\label{eq:HDG}
		-e^{2\phi} *_6 G ~& = ~  d\left[ -\frac{1}{Z_1}\, (du +\omega)\wedge (dv + \beta)\right] ~+~ \widehat G_1\,,
	\end{align}
\end{subequations}
where 
\begin{subequations}
	\begin{align}
		\widehat G_1 ~& \equiv ~*_4 \left(\cD Z_2 + \dot \beta Z_2 \right) + (dv+ \beta)\wedge \Theta^1\,,\\*
		\widehat G_2 ~& \equiv~ *_4 \left(\cD Z_1 + \dot \beta Z_1 \right) + (dv+ \beta)\wedge\left( \Theta^2+ \frac{Z_A^2}{Z_2^2}\, \cD\beta -2 \frac{Z_A}{Z_2}\, \cD \tA\right)\,.\label{eq:GHat2}
	\end{align}
\end{subequations}
The additional terms appearing in the definition of the $\Theta^2$ field now appear in the expression for $\widehat G_2$. 
It is important to note that despite these additional terms, due to the shift \eqref{eq:cFShift}, the expressions in \eqref{eq:GaugeNew} are consistent, assuming that the BPS equations are satisfied. 

The inclusion of terms containing the components of the U(1) gauge field in \eqref{eq:GHat2} is natural when writing the three-form fields in terms of the two-form potential $B$, as defined in \eqref{eq:FieldStrengthsDef}.
From the equations of motion and Bianchi identities for the three-form field strength, \eqref{eq:BosEOM}, we can see that while $G$ is not closed, the combination appearing on the left hand side of \eqref{eq:HDG} is. 
The latter can be thus locally written as
\begin{align}
	\label{eq:dBtildeDef}
	- e^{2\, \phi}*_6 G ~\equiv~ d \widetilde B\,.
\end{align}
Given the form of \eqref{eq:HDG} and the fact that, due to supersymmetry, all components of the fields are independent of the coordinate $u$, one can make the following Ansatz for the two-form potential
\begin{align}
	\label{eq:BtildeAns}
	\widetilde B ~& \equiv~ -\frac{1}{Z_1}\, (du +\omega)\wedge (dv + \beta)+ a_1 \wedge (dv + \beta) + \gamma_2\,.
\end{align}
where $a_1$ and $\gamma_2$ are respectively a one-form and a two-form on the base space.
This expression is consistent with \eqref{eq:dBtildeDef} provided these forms satisfy the constraints
\begin{subequations}
	\label{eq:Cond1}
	\begin{gather}
		\Theta^1 ~=~ \cD a_1 - \dot \beta\wedge a_1 + \dot \gamma_2\,,\\
		*_4\left(\cD Z_2 + Z_2\,\dot \beta\right) ~=~ \cD \gamma_2 - a_1 \wedge \cD\beta\,,
	\end{gather}
\end{subequations}
which are exactly analogous to those found in \cite{Bena:2011dd, Giusto:2013rxa}.
This should also be unsurprising, because acting on \eqref{eq:Cond1} with $\cD$ yields, after some algebra, the BPS equations for $Z_2$ and $\Theta^1$ which are unaltered by the presence of the U(1) gauge field. 

Despite $G$ not being closed, we can make the same reasoning as for $\widetilde B$ to argue for the absence of $du$ terms and make an Ansatz for the potential $B$ as
\begin{align}
	\label{eq:BDef}
	B ~& \equiv~ -\frac{1}{Z_2}\, (du +\omega)\wedge (dv + \beta)+ \left[ a_2 - \frac{Z_A}{Z_2} \, \tA\right] \wedge (dv + \beta) + \gamma_1\,,
\end{align}
where again $a_2$ and $\gamma_1$ are respectively a one-form and a two-form on the base space.
By using  \eqref{eq:FieldStrengthsDef} we find that this Ansatz is consistent with \eqref{eq:GaugeNew} provided 
\begin{subequations}
	\label{eq:Cond2}
	\begin{gather}
		 \Theta^2 ~=~ \cD a_2 - \dot \beta \wedge a_2+ \dot \gamma_1 + \dot \tA \wedge \tA\,,\\
		*_4\left(\cD Z_1 + Z_1\,\dot \beta\right) ~=~ \cD \gamma_1+ \cD \tA \wedge \tA  - a_2 \wedge \cD\beta\,,
	\end{gather}
\end{subequations}
which, upon the action with $\cD$,  reproduces the BPS equations \eqref{eq:Sub20}.
Note that by defining the $dv$  component of $B$ with a contribution from the U(1) fields, we are able to absorb several additional terms that are otherwise present in \eqref{eq:Cond2}.
The remaining quadratic terms are physical and cannot be reabsorbed by a redefinition of $\gamma_1$ -- they are responsible for the source terms in the third layer of the BPS equations \eqref{eq:Sub20}.

\subsection{Gauge invariance}

There are two types of gauge symmetries in this theory. 
Firstly, the usual diffeomorphism invariance, which is also present in the minimal gauge case and does not change with the addition of the U(1) gauge field. 
The Ansatz is invariant under shifts \cite{Gutowski:2003rg, Giusto:2013rxa, Bena:2015bea}
\begin{align}
	\label{eq:uvShift}
	v \to v + V(x^m)\,, \qquad u \to u + U(v, x^m)\,, 
\end{align}
and this leads to the transformations:
\begin{align}
	\label{eq:uvShiftAnsatz}
	\beta \to \beta - d_4 V\,, \qquad \omega \to \omega - \cD U\,, \qquad \cF \to \cF - 2 \dot U\,.
\end{align}

The three-form gauge field strength, and its Hodge dual, are invariant under the gauge transformations:
\begin{align}
	B \to B + d X\,, \qquad \widetilde B \to \widetilde B + d \widetilde X\,,
\end{align}
Because the supersymmetry requires $u$-independence, the potentials have very specific forms that we would like to preserve under the gauge transformations.  We therefore  require  $X$ and $\widetilde X$ to have  the form:
\begin{align}
	X ~\equiv~ X_v(v, x^n) \, (dv + \beta) + X_m(v, x^n)\, dx^m\,.
\end{align}
In practice this means that there always exists some freedom in choosing whether to describe the physical degrees of freedom present in the three-form field strengths as coming from one-forms $a_{1,2}$ or two-forms $\gamma_{1,2}$. 
On the other hand, the scalar functions $Z_1$ and $Z_2$ are invariant under this gauge transformation.

The U(1) gauge field has the standard gauge invariance  under an exact shift created by a scalar function $\lambda = \lambda(v, x^m)$:
\begin{align}
	\label{eq:U1Gauge}
	A \to A + d\lambda\,.
\end{align}
However, because $A$ also appears in the definition of the three-form strength, one needs to simultaneously shift the two-form $B$ by
\begin{align}
	\label{eq:U1GaugeB}
	B \to B - A \wedge  d\lambda\,,
\end{align}
to ensure gauge invariance of $G$.
Under this transformation, the components of the U(1) gauge potential change as
\begin{align}
	Z_A \to Z_A + Z_2\, \dot \lambda\,, \qquad \tA \to \tA - \cD \lambda\,,
\end{align}
where we assumed that $Z_2$ is invariant under this gauge symmetry. 
The associated change in $B$ induces
\begin{align}
	\label{eq:U1GaugeBComp}
	a_2 \to a_2 + 2\, \dot \lambda\, \tA-  \dot \lambda \, \cD \lambda\,, \qquad \gamma_1 \to \gamma_1 + \tA\wedge\cD\lambda\,,
\end{align}
with all other quantities remaining invariant. 
The quadratic dependence of $a_2$ on $\lambda$ is the consequence of the expression in the square parenthesis of \eqref{eq:BDef}.

Because the gauge transformation (\ref{eq:U1Gauge}) induces the transformation (\ref{eq:U1GaugeB}) in $B$, this results in a transformation of e $\Theta^2$ and $\cF$:
\begin{align}
	\label{eq:cFTh2GaugeTrans}
	\cF \to  \cF -2\, Z_A\, \dot\lambda - Z_2 \, \dot \lambda^2\,, \qquad  \Theta^2 \to \Theta^2 + 2 \, \dot \lambda\, \cD \tA + \dot\lambda^2\, \cD\beta\,.
\end{align}
Note that these changes are quadratic in $\lambda$.
It must be emphasized that \eqref{eq:cFTh2GaugeTrans} does not mean that any physical field is changed by the choice of gauge. 
Rather, the non-trivial mixing of degrees of freedom due to the shifts \eqref{eq:cFShift} and \eqref{eq:ThetaDefns} results in individual Ansatz quantities being gauge dependent. 

Thus it would seem that the first step in solving the BPS equations should be choosing an appropriate U(1) gauge that  fixes the Ansatz quantities.
Indeed, there are several choices of gauge which can simplify the BPS equations, for example, in \cite{Cariglia:2004kk} the authors chose $Z_A =0$, and this removes several terms in the BPS equations.
However,  for the purpose of finding smooth microstate geometries another choice might be more favourable, which is why we chose to undo the gauge-fixing at the cost of introducing redundancies in the description.

\subsection{Physical interpretation of the Ansatz and BPS equations}
\label{ssec:Shift}

Having  linearized the BPS system by introducing new fundamental fields into our Ansatz, we now take a step back and look at the physics of the fields that we have introduced. 

The six-dimensional  supergravity we are investigating is a compactification of  Type IIA supergravity on a $\IT^4$, or M-theory on a $\IT^5$.     Indeed, the degrees of freedom we retain in this compactification are those that ``have no legs'' on the $\IT^4$:  that is, only a scale factor and the volume form of the  $\IT^4$ appear in the Ansatz. 

The fields  $(Z_1,\Theta^2)$ and $(Z_2, \Theta^1)$,  respectively,  describe the behaviour of fundamental strings and NS5-branes, while $(Z_A,\tA)$ encode the RR fields stemming from  D0 and D4  sources of type IIA theory. The one form  $\beta$ encodes KKM fields and $\omega$ determines the  angular momentum.  The function, $\cF$, contains information about the sources of linear momentum along the $S^1(y)$ circle (\ref{yperiod}).

Compared to \cite{Bena:2011dd, Giusto:2013rxa}, the BPS equations analysed in this paper exhibit a novel feature in the first layer equations, breaking this system into three separate layers. 
This can be understood by observing that the three sets of equations describe objects that scale differently with the gravitational, or string, coupling.

The first set of equations determine $(Z_2, \Theta^1)$, and thus the NS5-branes, whose tension scales as $g_s^{-2}$, and are therefore the ``heaviest'' objects.
Next are the equations determining the   D-brane sources, whose tensions scale as   $g_s^{-1}$, and the BPS equations ties their behavior to that of the heavier NS5-branes. Specifically, 
this comes from the  factors of $Z_2$ and $\Theta^1$ in equation \eqref{eq:EqZA}.
Finally, the third set of equations involving $(Z_1, \Theta^2)$ describe  the lightest excitations, the fundamental strings, whose tension is independent of $g_s$. 
These are  influenced directly by the D-branes to which they couple, as can be seen in the source terms of \eqref{eq:Sub20}, and, as a result, these fields are influenced implicitly by the NS5 branes,

The fact that the three-form, $G$,  depends on the gauge dependent quantities $Z_A$ and $\tA$ arises from the fact that $G$ is not closed, due to the $F\wedge A$ terms in (\ref{eq:FieldStrengthsDef}).   However, in making the shift  \eqref{eq:cFShift}, we have also inserted the gauge dependent quantities into the metric, and this seems a little strange. 

An indirect way of seeing the necessity of this shift is to consider the system of equations studied in \cite{Giusto:2013rxa}, which is adapted to the study of the Type IIB system with D1-D5-P charges.
As described in Section \ref{ss:6Dtheory}, this can be described by $\cN = (1,0)$ supergravity theory coupled to two tensor multiplets. 
The D1-D5-P system can be dualised to the F1-NS5-P frame in Type IIA theory by first performing an S-duality followed by a T-duality along a compact direction. 
If we want to preserve the ``isotropy'' along the internal manifold, one has to T-dualise along either the $S^1(y)$ circle or the Gibbons-Hawking\footnote{Flat $\mathbb{R}^4$, in appropriate coordinates, can be written in Gibbons-Hawking form where the 4-dimensional space is decomposed into a circle fibered over a 3-dimensional base. More details are given in the following section.} fibre of the base space. 
Since such a transformation assumes that the geometry is independent of the coordinate along which we dualise, a generic supergravity solution in the D1-D5-P system, which depends on both coordinates,  does not have a valid description within the F1-NS5-P frame in Type IIA supergravity.%
\footnote{Any dependence on the coordinate along which we T-dualise generates a tower of massive Kaluza-Klein modes, which takes us out of the regime of validity of supergravity.}
%
Despite this, such a duality chain allows us to correctly identify how different Ansatz quantities in the F1-NS5-P system are associated with the brane sources and one finds that the shifts are necessary in order to avoid mixing of degrees of freedom. 

From a physical point of view, the two contributions to the $g_{vv}$ component of the metric can be explained by a ``dual'' role of  U(1) charges along the $S^1(y)$ circle. 
Any source that moves along this circle and carries momentum will contribute to $\cF$ with a minus sign, which corresponds to a positive contribution to the momentum charge.  These contributions come from the source terms in (\ref{eq:Lay2eq2New}).
On the other hand, the $Z_A^2/Z_2$ term points in the opposite direction and thus reduces momentum. 
One can interpret this as a version of Faraday's law of induction in which the moving  U(1) sources induce a response ``slowing down'' the geometry.

Finally, there is the  approximate symmetry between the pairs $(Z_A, \tA)$ and $( \cF, \beta)$  that is apparent in the BPS equations.
This can be easily understood from the perspective of M-theory, where the fields $(Z_A, \tA)$ describe the fibration in which the extra spatial circle, described by $x^{10}$, is compactified to give the IIA theory.
The quantities $\cF$ and $\beta$ have the same roles but for the $S^1(y)$ circle, as can be seen  from the metric Ansatz \eqref{eq:MetAnsNew}.
Thus, from the M-theory perspective, one would expect  an $SL(2,\ZZ)$ symmetry acting on $S^1(x^{10})\times S^1(y)$ and thus upon the fields $(Z_A, \tA)$ and $( \cF, \beta)$.

However, we have explicitly broken this symmetry because all our fields are independent of $x^{10}$, but are allowed  to depend on $v$.  Moreover, there is also the asymmetry coming from the fact that  $\tA$ contributes to a vector multiplet while $\beta$ contributes to a tensor multiplet: this results in slightly different interactions with the other fields. Indeed, 
we will see this difference in Section \ref{5Dsolutions}, when we  analyse solutions that are independent of $v$, and compactify to five dimensions. All these fields become vector multiplets in five dimensions, but the intersection form that governs the interactions of these vector multiplets has a slightly more subtle symmetry involving $(Z_A, \tA)$ and $( \cF, \beta)$.    

\section{Examples}
\label{sec:Examples}

In this section we examine several examples of explicit solutions to the BPS equations.  

We start with the simplest families of solutions:  those that are $v$-independent, and  can therefore  be compactified to five dimensions.    The tensor and vector multiplets of the six-dimensional theory all reduce to five-dimensional vector multiplets, putting all of these fields on the same footing.  While this leads to a relatively straightforward, five-dimensional BPS system, it reveals how the degrees of freedom of the six-dimensional vector multiplet interact with the other fields of the six-dimensional supergravity.

The other examples are singular brane configurations with non-trivial $v$-dependence, and based on the results of \cite{Bena:2022sge}.  These solutions, which are obtained primarily through dualities,  provide non-trivial tests of the BPS equations and illustrate some of the physical effects described in Section \ref{ssec:Shift}.

\subsection{Five-dimensional solutions}
\label{5Dsolutions} 

\subsubsection{The five-dimensional theory}

The base metric (\ref{basemet}) must be hyper-K\"ahler, and for simplicity we take it to be a  Gibbons-Hawking (GH) space \cite{Gibbons:1978tef}.  One should, of course, remember that this metric is allowed to be ambi-polar \cite{Giusto:2004kj,Bena:2005va,Berglund:2005vb, Bena:2007kg,Warner:2019jll}.  We therefore have:
\begin{align}
	\label{eq:GHForm}
	ds_4^2  ~&=~ \frac{1}{V}(d\psi + A_{GH})^2 ~+~ V\, dy^a\, \,dy^a\,,
\end{align}
where the fibre is taken to be periodic with $\psi \sim \psi + 4\pi$. 
The three-dimensional space described by the $y^a$ is  flat with $A_{GH}$ and  $V$ satisfying:
\begin{align}
	\label{eq:GHCond}
	*_3 dA_{GH} ~=~ dV\,,
\end{align}
from which it  follows that 
\begin{align}
	\label{eq:GHCond2}
	*d*dV~=~0\,, \qquad d*dA_{GH} ~=~ 0\,.
\end{align}

We also restrict our focus to solutions that are independent of the GH fibre $\psi$.
As a result, all quantities are functions of only the $y^a$.
Firstly, and for this section alone, we define 
\begin{align}
	\label{eq:GHIdent}
	Z_3  ~\equiv~ - \widetilde \cF\,, \qquad \Theta^3 ~\equiv ~d\beta\,,\qquad 
	Z_4  ~\equiv~{\sqrt2}\, Z_A \,, \qquad \Theta^4 ~\equiv~ - {\sqrt2}\, d\tilde A\,.
\end{align}
The other $2$-forms are closed and have simple potentials:
\begin{align}
	\label{eq:GHThetas1}
	\Theta^1 ~=~ d a_1\,, \qquad \Theta^2 ~=~ d a_2\,.
\end{align}
One should note that the simple ``decoupled form'' of   $\Theta^2$ is a direct consequence of the definition \eqref{eq:BDef} and the inclusion of the additional term  in the square parentheses multiplying the $(dv + \beta)$ component.

As usual, one decomposes all the forms in terms of components along the GH fibre.
For one-forms this means
\begin{subequations}
	\label{eq:OneFormDec}
	\begin{align}
		a_1~ & =~ \frac{K^1}{V}(d\psi + A_{GH}) + \sig{1}\,,\\
		a_2 ~& =~ \frac{K^2}{V}(d\psi + A_{GH}) + \sig{2}\,,\\
		\beta~ & = ~\frac{K^3}{V} (d\psi + A_{GH}) + \sig{3}\,,\\
		\tilde A~ & =~-\frac{1}{\sqrt{2}} \left[\frac{K^4}{V}(d\psi + A_{GH}) + \sig{4}\right]\,,\\
		\omega ~& =~ \mu\,(d\psi + A_{GH}) + \bar \omega\,,
	\end{align}
\end{subequations}
where $\sig{I}$ and $\bar \omega$ are one-forms on the three-dimensional flat space and $\mu$, and the $K^I$ are scalar functions. 

The next  layers can be slightly reorganised, as all $Z_I$ and $\Theta^I$ are now on the same footing.
Firstly, all two-forms are self-dual and closed on the base space
\begin{align}
	\label{eq:BubbleEq1}
	*_4 \Theta^I ~=~ \Theta^I\,, \qquad d_4 \Theta^I ~=~ 0\,.
\end{align}
The equations that link  $Z_I$ to  $\Theta^I$ are simply Laplace equations on the GH space with $\Theta^I$ acting as sources
\begin{subequations}
	\label{eq:BubbleEq2}
	\begin{align}
		\Box_4 Z_1 ~&=~  *_4 \Theta^2 \wedge \Theta^3 -\frac12 *_4 \Theta^4 \wedge \Theta^4\,,\\
		\Box_4 Z_2 ~& =~  *_4 \Theta^1 \wedge \Theta^3\,,\\
		\Box_4 Z_3 ~& =~  *_4 \Theta^1 \wedge  \Theta^2\,,\\
		\Box_4 Z_4 ~&=~ -  *_4 \Theta^4 \wedge \Theta^1\,.
	\end{align}
\end{subequations}
Finally, the BPS equation dealing with the one-form $\omega$ is given by
\begin{align}
	\label{eq:BubbleEq3}
	d_4\omega + *_4d_4 \omega ~=~  Z_1\, \Theta^1 + Z_2\,\Theta^2 + Z_3 \,\Theta^3 +  Z_4\, \Theta^4 \,.
\end{align}
These are the well-known, linearized five-dimensional BPS equations \cite{Bena:2004de,Bena:2005va, Bena:2007kg}. 

In general, such BPS systems \cite{Gauntlett:2003fk,Gutowski:2004yv,Bena:2004de}  are characterized by the five-dimensional structure constants,  $C_{IJK}$:
\begin{subequations}
	\begin{gather}
		*_4 \Theta^I ~=~ \Theta^I\,, \qquad d\Theta^I ~=~ 0\,,\qquad  \Box_4 Z_I ~=~  \frac12 C_{IJK}\,*_4\Theta^J \wedge \Theta^K\,,\\
		d_4 \omega + *_4 d\omega ~= ~ Z_I\, \Theta^I\,,
	\end{gather}
\end{subequations}
Our system therefore has:
\begin{align}
	C_{123} ~=~1 \,, \qquad C_{144} ~=~ -1\,.
	\label{OurCIJK}
\end{align} 
As we remarked in Section \ref{ssec:Shift}, this interaction does not have a symmetry that interchanges  $(Z_A, \tA)$ and $( \cF, \beta)$, which is now equivalent to $3 \leftrightarrow 4$.  Interestingly enough, if one defines 
\begin{equation}
	Z_\pm ~\equiv~\frac{1}{2}\, (Z_3  \pm Z_2)  \,, \qquad \Theta^\pm ~\equiv~\frac{1}{2}\,( \Theta^3  \pm  \Theta^2)  \,.
\end{equation} 
then the structure constants become 
\begin{align}
	C_{1++} ~=~1 \,, \qquad C_{1--}~ =~ -1 \,,  \qquad C_{144} ~=~ -1\,.
\end{align} 
and there is a symmetry with $- \leftrightarrow 4$.  Whether this can be uplifted to  realize an equivalence between the M-theory circle and the $y$-circle is unclear.  Indeed, we suspect that such an equivalence will require adding more degrees of freedom to the six-dimensional supergravity

For comparison, we note that the   corresponding system for the D1-D5-P system, in which the momentum carriers are encoded in a six-dimensional tensor multiplet, reduces to a five-dimensional supergravity with 
\begin{align}
	C_{123} ~=~1 \,, \qquad C_{344} ~=~ -1\,.
\end{align} 
%

\subsubsection{Solutions of the five-dimensional theory}

To write the solutions we introduce an orthonormal frame for the GH space
\begin{align}
	\label{eq:GBFrame}
	\hat e^1 ~=~ \frac{1}{\sqrt V}\, (d\psi + A_{GH}) \,,\qquad\qquad   \hat e^{a+1} ~=~ \sqrt V\,  \tilde e^a\,, \quad a = 1,2,3\,,
\end{align}
where $\tilde e^a$ are the frames on the three-dimensional flat space. 
Using these, one can then construct (anti-)self-dual two-forms \cite{Bena:2008wt}
\begin{align}
	\label{eq:GHASDTwoForms}
	\Omega^{a}_{\pm} & ~\equiv~ \hat e^1 \wedge \hat e^{a+1} ~\pm~ \frac12\,\epsilon_{abc}\, \hat e^{b+1} \wedge \hat e^{c+1}\,.
\end{align}
The set of anti-self-dual two forms $\Omega_-^{a}$ are precisely the three two-forms, $J^A$, that define the hyper-K\"ahler structure.  

By inserting the decomposition \eqref{eq:OneFormDec} into the expressions for the two-forms $\Theta^I$, one finds that the self-duality condition is satisfied if
\begin{align}
	*_3dK^I ~=~ -d\sig{I}\,,
\end{align}
from which it follows that $K^I$ have to be harmonic on the three-dimensional base:%
\footnote{For $\psi$-independent solutions, the four-dimensional Laplace operator is given by $\Box_4 = V^{-1}\, \Box_3$. 
	Hence a function that is harmonic on the three-dimensional space is also harmonic on the full four-dimensional Gibbons-Hawking space.} 
\begin{align}
	\Box_3 K^I ~=~ 0\,.
\end{align}
Furthermore,  $\Theta^I$ can be rewritten in the basis of the three self-dual two forms $\Omega^a_+$ as
\begin{align}
	\Theta^I  ~=~ - \sum_{a=1}^3\,\pd_a\left(\frac{K^I}{V}\right)\,\Omega_+^a\,.
\end{align} 
Inserting these expressions into \eqref{eq:BubbleEq2} one finds that the solutions for $Z_I$ are given by
\begin{subequations}
	\label{eq:BubbleSol1}
	\begin{align}
		Z_1 ~&=~ L_1 + \frac{K^2 \, K^3 - \frac12\, \left(K^4\right)^2}{V}\,,\\
		Z_2 ~& =~ L_2 + \frac{K^1\, K^3}{V}\,,\\
		Z_3 ~&=~ L_3 + \frac{K^1\, K^2}{V}\,,\\
		Z_4 ~&=~ L_4 - \frac{K^1\, K^4}{V}\,,
	\end{align}
\end{subequations}
where $L_I$ are arbitrary harmonic functions. 
The equation determining $\omega$ is separated into two parts which determine $\mu$ and $\bar \omega$ in terms of known sources
\begin{subequations}
	\begin{align}
		\Box_3 \mu ~&=~ \frac{1}{V} *_3 d_3*_3\left[ V\, Z_I\,d\left(\frac{K^I}{V}\right)\right]\,,\\
		*_3 d_3\bar \omega ~& =~ V \, d\mu - \mu\, dV - V\, Z_I\,d \left(\frac{K^I}{V}\right)\,.
	\end{align}
\end{subequations} 
The first equation is solved  by
\begin{align}
	\mu ~=~ \frac{M}{2} + \frac{K^1\, L_1 + K^2\, L_2 +K^3\, L_3 +K^4\, L_4 }{2\,V}+ \frac{K^1\, K^2\,K^3 - \frac12\,K^1\,\left(K^4\right)^2}{V^2}\,,
\end{align}
with $M$ being a freely choosable harmonic function. 
Inserting this expression into the lower equation yields 
\begin{align}
	\label{eq:BarOmegaEq}
	*_3 d_3\bar \omega ~& =~ \frac12\, V \, dM  - \frac12\,M\, dV+ \frac12\left( K^I\, dL_I - L_I\, dK^I\right)\,,
\end{align}
much as in \cite{Bena:2007kg}, only that the implicit sum is over four different harmonic functions.

The solution can be written more compactly in terms of the structure constants:
\begin{align}
	Z_I ~=~ L_I + \frac12\,C_{IJK}\frac{K^J\,K^K}{V}\,,\qquad 
	\mu ~=~ \frac{M}{2}+ \frac12\,\frac{L_I\,K^I}{ V}+ \frac1{3!}\,C_{IJK}\frac{K^I\,K^J\,K^K}{V^2}\,,
\end{align}
together with \eqref{eq:BarOmegaEq}.

\subsection{Fundamental string with momentum charge}
\label{ssec:F1P}

To construct non-trivial $v$-dependent brane configurations, we start with a fundamental string carrying momentum.  This solution is elementary with fields sourced by harmonic functions.  We will then combine this with another  simple solution obtained in the next section, to construct a non-trivial hybrid solution in Section \ref{ss:hybrid}.  

We take the spacetime to be $\IR^{4,1}\times S^1\times T^4$, with the string wrapped along the $S^1$ circle, which is also the direction of the linear momentum. 
We allow the string to describe an arbitrary curve in the four-dimensional base space, given by a set of profile functions $g_m(v)$, while smearing the string along the  $T^4$.
The solution is thus independent of the $T^4$ and can be described in a six-dimensional theory of gravity with the bosonic fields given by \cite{Dabholkar:1995nc, Callan:1995hn}%
\footnote{Alternatively one can work directly in a six-dimensional setting.}
	\begin{align}
		ds^2 ~&=~ -\frac2{\sqrt{H_1}} \, dv\left[ du- (1-H_1)\dot g_m(v) \, dx^m + \frac12 (1-H_1)\dot g_m^2(v) \, dv\right] +\sqrt{H_1}\, \delta_{mn}\, dx^m\, dx^n\,,\nonumber\\
		F~& =~ 0\,,\qquad G ~=~ *_4 dH_1+ dv \wedge *_4 d\left[  -(1- H_1)\,\dot g_m(v)\, dx^m\right]\,,\qquad 
		e^{2 \phi} ~=~ \frac{1}{H_1}\,,\label{eq:F1PSolProfR4}
	\end{align}
where
\begin{align}
	H_1 ~\equiv~ 1 + \frac{Q_1}{\left|x_m - g_m(v)\right|^2}\,,
\end{align}
is a harmonic function and $Q_1$ is the supergravity charge of the fundamental string. 

The solution lies in the pure NS sector solution in Type II supergravity and the Maxwell field strength, $F$, is trivial.   
One can also calculate
\begin{align}
	- e^{2\phi}*_6 G ~&=~  d\left[-\frac{1}{H_1}\left(du-(1- H_1)\,\dot g_m(v)\, dx^m\right) \wedge dv \right] \,.
\end{align}

By comparing this solution with the supersymmetric Ansatz, one is able to read off the following quantities 
\begin{equation}
	\begin{split}
		\label{eq:F1PAns}
		h_{mn}& = \delta_{mn}\, \qquad Z_1 = H_1\,, \qquad Z_2 = 1\,,\qquad  \cF  = \left(1-H_1\right) \,\dot g_m^2(v)\,\\
		Z_A &= \tA = \beta =\psi=0\,,\qquad 
		\omega  = -\left(1-H_1\right)\,\dot g_m(v) \, dx^m\,,\\
		\Theta^1 &= 0\,, \qquad 	\Theta^2 = \left(1+*_4\right)d_4\left(- \left(1-H_1\right)\, \dot g_m(v)\, dx^m\right) = \left(1+*_4\right)d_4 \omega\,,
	\end{split}
\end{equation}
which can be shown to solve the BPS equations. 
Out of all the quantities in the first layer of the  BPS equations,  only the pair $(Z_1, \Theta^2)$ is excited, showing that these quantities describe the electric and magnetic properties of fundamental strings.

This solution is related, via a T-duality along one of the trivial directions of the internal manifold followed by S-duality, to the D1-P spiral solution studied in \cite{Bena:2011dd}.

\subsection{NS5-P brane configuration with local D0-D4 charges}
\label{ssec:NS5P}

 Starting from the ten-dimensional F1-P solution, one can use a chain of S and T-dualities to obtain a solution with NS5-branes carrying momentum via D0-D4 brane charges localised on its world-volume \cite{Bena:2022sge}.  This solution, while still based on harmonic sources, involves non-trivial Maxwell fields sourced by the D0-D4 branes.

Working again in a spacetime that is asymptotically $\IR^{4,1}\times S^1\times T^4$, and smearing the  D0-charges along the $T^4$, the solution (see equation~(3.8) of \cite{Bena:2022sge}) can be reduced to six-dimensions.  We generalise the solution of \cite{Bena:2022sge} by letting the NS5-brane distribution in the base space be given by a generic profile described by a vector, $g_m(v)$, in $\IR^4$.
This solution also has another  functional degree of freedom, $F(v)$,  describing  the D0-D4 charge distribution along the $S^1$ circle.  This function must be periodic in the manner determined by (\ref{uvtyreln}) and (\ref{yperiod}):  $F(v) = F(v + \sqrt2\, \pi\, R_y)$.

The bosonic fields describing such a configuration are given by
\begin{align}
		ds_{6}^2 &= - 2 \frac{dv}{\sqrt{H_5}}\, \left[ du- \left(1-{H_5}\right)\, \dot g_m(v)\, dx^m +\frac12\left(\dot g_m^2(v)\left(1-H_5\right)- \dot F(v)^2\,\left(1- \frac1{H_5}\right)\right) dv \right]\nonumber \\
		&\quad + \sqrt {H_5} \, \delta_{mn}\, dx^m\, dx^n\,,\nonumber\\
		F~&=~ dv\wedge d_4\left[ \dot F(v)\left(1- \frac1{H_5}\right)\right]\,,\\
		G~&  =~ - d\left[-\frac{1}{H_5} (du - \left(1-H_5\right)\, \dot g_m(v)\, dx^m) \wedge dv\right]\,, \qquad e^{2\phi} ~=~ H_5\,,	\nonumber
\end{align}
where we take $H_5$ to be the harmonic function sourced by the profile on the base space:
\begin{align}
	\label{eq:HarFun2}
	H_5 ~\equiv~ 1 + \frac{Q_5}{\left|x_m - g_m(v)\right|^2}\,,
\end{align}
where $Q_5$ denotes the supergravity charge associated with NS5-branes. 
As before, one can calculate
\begin{align}
	- e^{2\phi}*_6 G &~=~ *_4 dH_5 + dv \wedge *_4d\left[-(1-H_5)\, \dot g_m(v)\,dx^m\right]\,,
\end{align}
and read off the Ansatz quantities:
\begin{align}
	\label{eq:AnsatzNS5P2}
	h_{mn}& = \delta_{mn}\, \qquad
	Z_1 = 1\,, \qquad Z_2 = H_5\,, \qquad 
	\beta = \psi=0\,,\qquad \omega = - \left(1-H_5\right)\, \dot g_m(v)\, dx^m\,,\nonumber\\
	\Theta^1 &= \left(1+*_4\right)d_4\left(- \left(1-H_5\right)\, \dot g_m(v)\, dx^m\right) = \left(1+*_4\right)d_4 \omega\,, \qquad \Theta^2 = 0\,.
\end{align}

We note that in this solution $(Z_1, \Theta^2)$  are trivial while  $(Z_2, \Theta^1)$  are now excited, highlighting the fact that these degrees of freedom capture the fundamental string and NS5 brane excitations respectively.

By comparing the expression of the U(1) gauge field with the general Ansatz \eqref{eq:Fform} and \eqref{eq:OmegaTildeF}, we find that one needs to solve:
\begin{align}
	d_4 \tA ~=~ 0 \,, \qquad - \dot \tA - d_4\left(\frac{Z_A}{H_5}\right) ~=~ d_4 \left[ \dot F(v)\left(1- \frac1{H_5}\right)\right]\,.
\end{align}
The solutions are given
\begin{align}
	\label{eq:ExampleU1GaugeChoice}
	Z_A ~=~ - c_A\, \dot F(v)\left(H_5-1\right)\,, \qquad \tA ~=~ - \left(1-c_A\right) \, F(v) \, d_4 \left(1-\frac1{H_5}\right)\,,
\end{align}
where $c_A$ is a free parameter corresponding to a U(1) gauge freedom.
The choice  influences the expression for $\cF$, which is given by:
\begin{align}
	\cF ~=~ \dot g_m^2(v) \left(1-H_5\right) + \dot F^2(v)\left[2\,c_A^2-1 - c_A^2\, H_5 + \frac{1}{H_5}\left(1-c_A^2\right)\right]\,,
\end{align}
but note that the BPS equations are satisfied regardless of the value of $c_A$.  Also note that $c_A$ cancels in the metric Ansatz (\ref{eq:MetAnsNew}).

The two interesting choices are $c_A =0$ and $c_A = 1$, in which either $Z_A$ or $\tA$ vanish. 
If  $Z_A$ vanishes, both $\cF$ and $\tA$ contain factors of $(1- H_5^{-1})$ which are not harmonic on the base space.  
One may expect that finding more complicated solutions in such a gauge might be a little more technically challenging. 
Taking $c_A = 1$, with $\tA = 0$, achieves the opposite:   $Z_A$ and $\cF$ are harmonic on the base space, from which we can extrapolate that some version of this gauge choice  might be more favourable in the explicit construction of non-trivial solutions. 
%

\subsection{F1-NS5-P solution with  local D0-D4 charges}
\label{ss:hybrid}

If one considers the NS5-P solution presented above, but puts all the brane sources at a single point in the base space, then one is able to use S and T-dualities and the existing BPS Ansatz in the D1-D5-P system \cite{Giusto:2013rxa} to add fundamental strings to the brane configuration and create a three charge solution \cite{Bena:2022sge}.
One of the interesting properties of this solution is that, even though it has three asymptotic charges and thus looks like a three charge black hole, its horizon area vanishes. 

With our BPS Ansatz, we can analyse this solution in more detail.
We show that the absence of a finite-size horizon is the consequence of delicate cancellations between the two contributions to the $g_{vv}$ component of the metric, which ultimately determines the size of the $S^1_y$ circle at the horizon. 
Furthermore, using a simplified model, we show how the U(1) field directly determines the near-horizon behaviour of such solutions and comment on possible extensions. 

We begin by writing the solution given in equation (3.12) of \cite{Bena:2022sge} in the corresponding six-dimensional form
\begin{align}
		ds^2 ~&=~ - \frac{2}{\sqrt{H_1\, H_5}}\, dv\, \left[ du - \frac{\dot F(v)^2}{2}\,\left(1- \frac1{H_5}\right) \,dv\right] + \sqrt{H_1\, H_5} \,\delta_{mn} \,dx^m\, dx^n\,,\nonumber\\*
		F ~&=~ dv\wedge d_4\left[ \dot F(v)\left(1- \frac1{H_5}\right)\right]\,,	\label{eq:NS5F1P-D0D4}\\*	
		G ~&=~d\left[- \frac{1}{H_5}\, du\wedge dv\right] + *_4 dH_1\,, \qquad e^{2\phi} ~=~ \frac{H_5}{H_1}\,,\nonumber
\end{align}
where, because the sources are localised at a point in $\IR^4$,   the harmonic functions are
\begin{align}
	\label{eq:HarmFunPoint}
	H_1 ~=~ 1 + \frac{Q_1}{r^2}\,, \qquad H_5 ~=~ 1 + \frac{Q_5}{r^2}\,.
\end{align}
As before $F(v)$ is a periodic function describing the charge distribution of D0-D4 branes along the NS5 world-volume. We can also calculate
\begin{align}
	- e^{2\phi}*_6 G ~=~ d\left[- \frac{1}{H_1}\, du\wedge dv\right] + *_4 dH_5\,.
\end{align}
Working in the gauge $\tA = 0$ (see equation~\eqref{eq:ExampleU1GaugeChoice}), we can read off the corresponding Ansatz quantities%
\footnote{There are other Ansatz quantities that are non-vanishing in order to satisfy the conditions \eqref{eq:Cond1} and \eqref{eq:Cond2}, but these can be trivially determined from the quantities listed in \eqref{eq:F1NS5P-Ansatz}.}
\begin{equation}
	\begin{split}
		Z_1 = H_1\,, \quad Z_2 = H_5\,, \quad Z_A = - \dot F(v)\left(H_5 -1\right)\,, \quad \cF = - \dot F^2(v)\left(H_5 -1\right)\,,\\
		h_{mn} = \delta_{mn}\,, \qquad \omega = \beta = \tA =0\,,\qquad \gamma_2 = \gamma\,,\qquad  \Theta^1 =  \Theta^2 =\psi = 0\,,	\label{eq:F1NS5P-Ansatz}
	\end{split}
\end{equation}
where $\gamma$ is a two-form defined through $d \gamma \equiv *_4 dH_5$.
After substituting \eqref{eq:HarmFunPoint}  into the expressions for $\cF$ and $Z_A$ gives
\begin{align}
	\label{eq:Ftilde_and_Z_A}
	\cF ~=~ - \dot F^2(v)\, \frac{Q_5}{r^2}\,, \qquad Z_A ~=~ - \dot F(v)\, \frac{Q_5}{r^2}\,.
\end{align}
This explicitly shows that all scalar functions in the solution \eqref{eq:F1NS5P-Ansatz} are given by harmonic functions with a (possibly $v$-dependent) point-like source.
Indeed, one can show that this is a  solution to the BPS equations and the bosonic equations of motion.

Asymptotic analysis of this solution shows that it contains three global charges, corresponding to fundamental strings, NS5-branes, and momentum. 
On the other hand, the near-brane limit ($r\to 0$) shows that there is no finite-sized horizon \cite{Bena:2022sge}.  
With the BPS Ansatz we can offer an additional explanation. 
The $g_{vv}$ component of the metric contains two factors, coming from the Ansatz quantities $\cF$ and $Z_A$.  
As already discussed, these contribute to the momentum charge with opposite signs. 
At a generic radial distance the magnitudes of the two terms will be different, however, at $r=0$ these terms precisely cancel out
\begin{align}
	\label{eq:gvvExp}
	g_{vv}\sim	\cF + \frac{Z_A^2}{Z_2} = - \dot F^2(v)\, \frac{Q_5}{r^2} + \frac{\left(- \dot F(v)\, \frac{Q_5}{r^2}\right)^2}{1 + \frac{Q_5}{r^2}} \xrightarrow{r\to 0} - \dot F^2(v)\, \frac{Q_5}{r^2} +  \dot F^2(v)\, \frac{Q_5}{r^2}\Big[1 + \cO\left(r^2\right) \Big]\,.
\end{align}
Near the brane sources the ``standard'' momentum due to the longitudinal D0-D4 waves is cancelled out by the opposing effect coming from the gauge field, sourced by the same charges. 

The subleading terms in the expansion \eqref{eq:gvvExp} become important when taking the near-horizon limit \cite{Maldacena:1997re}, which can be in practice implemented by taking the limit 
\begin{align}
	\label{eq:SmallrExp}
	r^2 \ll Q_1, Q_5\,,
\end{align}
and retaining the leading order terms.%
\footnote{In \cite{Bena:2022sge} it was shown that taking the near-horizon this way is equivalent to the scaling of string theory parameters described in \cite{Maldacena:1997re}.
However, due to the presence of $H_5$ in \eqref{eq:gvvExp}, one cannot simply `drop the $1$'s', as this results in the loss of the leading order term of the $g_{vv}$ component.}
The metric is to leading order given by
\begin{align}
	\label{eq:near-horizon_IIA}
	ds^2 ~=~ - \frac{2\, r^2}{\sqrt{Q_1 \, Q_5}}\, dv \left( du - \frac{\dot F^2(v)}{2} dv\right) + \frac{\sqrt{Q_1 \, Q_5}}{r^2}\, dr^2 + \sqrt{Q_1 \, Q_5}\, d\Omega_3^2\,,
\end{align}
where we have used spherical coordinates for the base space metric in which case \mbox{$ds_4^2 = dr^2 + r^2 \, d\Omega_3^2$}. 
This metric is locally  AdS$_3 \times S^3$ with radius $R_{\rm AdS}^2 \equiv \sqrt{Q_1 \, Q_5}$, which can be seen perhaps more clearly by introducing standard spacetime coordinates $(t,y)$ defined in (\ref{uvtyreln}).

The additional term in the metric, which represents a deviation from the usual AdS$_3 \times S^3$ expression, does not affect the local behaviour of the geometry. 
This can be also seen by examining the gauge fields, where we find that the U(1) field becomes pure gauge, while the three-form field strength decomposes, to leading order, as 
\begin{align}
	\label{eq:AdSG}
	G ~=~ -\frac{2}{Q_5} \,{\rm Vol}_{{\rm AdS}_3}~+~ 2\, Q_1\, {\rm Vol}_{S^3}\,,
\end{align}
where ${\rm Vol}_{{\rm AdS}_3}$ and ${\rm Vol}_{S^3}$ are respectively the volume forms of AdS$_3$ and $S^3$ with unit radius. 
Such a form for $G$ is indicative of empty AdS$_3 \times S^3$.
However, the additional, $F(v)$-dependent term in the metric does not vanish at asymptotic infinity. It thus represents a non-trivial deformation of the metric at the conformal boundary.
The interpretation of this result is that the solution \eqref{eq:NS5F1P-D0D4} does not fit into an AdS region as the momentum is localised in the transition zone between the AdS and flat regions.

\subsubsection{A toy example}

To further illustrate the balance of momentum charges, we can consider a related, but simplified example.
If only scalar Ansatz quantities are excited on a base space with a flat metric, as in \eqref{eq:F1NS5P-Ansatz}, then the BPS equations reduce to a standard harmonic Ansatz in four-dimensional Euclidean spacetime. 
Therefore,  the following choice of functions is a solution to the BPS equations
\begin{align}
	\label{eq:NaiveSolution4Charges}
	Z_1 = H_1\,, \qquad Z_2 = H_5\,, \qquad Z_A = \frac{\sqrt2\,Q_A}{r^2}\,, \qquad \cF = - 2\, \frac{Q_P}{r^2}\,,
\end{align}
where $H_1$ and $H_5$ are given in \eqref{eq:HarmFunPoint}.
The important difference between this solution and \eqref{eq:F1NS5P-Ansatz} is that the charges associated to $\cF$ and $Z_A$ fields are now independent constants, $Q_P$ and $Q_A$.
The former corresponds to the momentum charge measured at infinity as $Q_A$ contributes only at subleading orders in the large $r$ expansion \cite{Myers:1986un}
\begin{align}
	\label{eq:AsyExp}
	\cF + \frac{Z_A^2}{Z_2}~ \xrightarrow{r\to \infty}~ - \frac{2 Q_P}{r^2} + \frac{2Q_A^2}{r^4} + \cO\left(r^{-6}\right)\,.
\end{align}
However, these two terms compete in the interior of the geometry, which is most significant in determining the size of the $S^1$ circle. 
Using the  coordinates (\ref{uvtyreln}), one finds that the metric coefficient of $dy^2$ as a function of the radial coordinate is given by
\begin{align}
	g_{yy}(r) ~=~ \frac{ \left(Q_5 + r^2\right)\left(Q_P + r^2\right)- Q_A^2}{ \left(Q_5 + r^2\right)^{\frac32}\,\left(Q_1 + r^2\right)^{\frac12}}\,.
\end{align}
Since all the charges are positive, we can focus on the numerator because  the denominator is always positive.

If $Q_A$ is too large, then $g_{yy}$ can vanish or become negative, which means there are closed timelike curves (CTCs).  Indeed, 
since the numerator is  monotonically increasing as a function of $r$, for $r \geq 0$, it is sufficient to analyse its behaviour at $r = 0$ where
\begin{align}
	\label{eq:gyyExp}
	 g_{yy}(r=0) ~=~ \frac{ \,Q_5\, Q_P - Q_A^2}{ \sqrt{Q_1\, Q_5^{3}}}\,.
\end{align}
It follows that one must take
\begin{align}
	\label{eq:BoundQA}
	 Q_A^2 ~\leq~ \, Q_5\, Q_P\,,
\end{align}
to avoid CTC's. 

Moreover, $g_{yy}$  is also a factor  in the expression for the horizon area,   the quadratic combination of charges in the numerator explicitly determines the size of the horizon:
\begin{align}
	\label{eq:HorizonArea}
	A_H ~=~ 4 \,\pi^3\,R_y\,\sqrt{Q_1\,\left(Q_5\, Q_P-Q_A^2\right)}\,.
\end{align}
This is precisely what one would expect from the structure constants, $C_{IJK}$, given in (\ref{OurCIJK}).  Indeed, from the five-dimensional perspective, any such BPS black hole can be generated by acting on a three-charge seed solution with classical duality symmetries \cite{Cvetic:1996zq} and the horizon area can be reduced to the standard three-charge form with \mbox{$A_H \sim \sqrt{Q_1\, Q_+\, Q_-}$}, where
\begin{equation}
	Q_{\pm} ~\equiv~ \sqrt{Q_5\, Q_P} ~\pm~ |Q_A|\,.
\end{equation}

\begin{figure}[t!]
	\centering
 	\includegraphics[width=\linewidth]{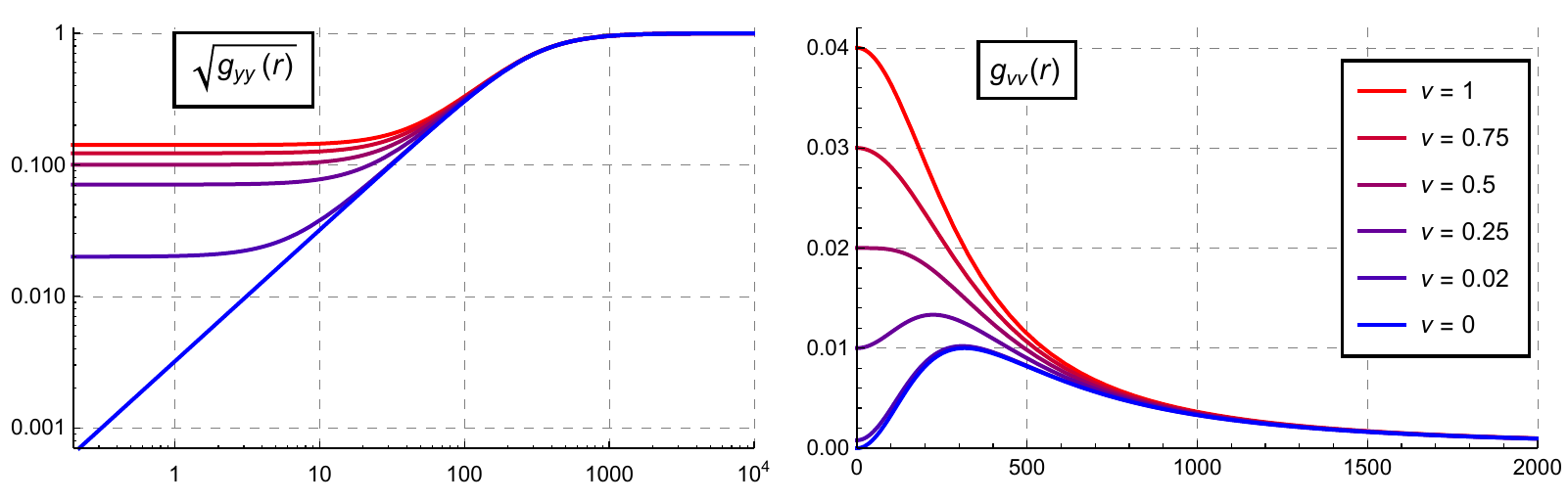}
	\caption{Left: The log-log plot of $\sqrt{g_{yy}}$, which determines the proper length of the $S^1(y)$ circle,  as a function of the radial coordinate for different values of $\nu$, defined in \eqref{eq:nuDef}.
		For $\nu =0$ the $y$-circle pinches off at $r=0$, followed by a region of linear increase indicating an AdS$_3$ spacetime.
		At large values of $r$ the circle stabilises signalling the onset of the asymptotically flat region. 
		For $\nu \neq 0$ there is another region where the size of the $S^1(y)$-circle is constant due to the  AdS$_2\times S^1$ throat region. 
		With increasing $\nu$ the transition between the AdS$_3$ and the throat region appears at larger values of $r$.
		Right: Plot of $g_{vv}(r)$, which contains the information about the momentum of the system,  for various values of $\nu$, while keeping $Q_P$ fixed to ensure equal asymptotic behaviour.
                 For $\nu =1$ the source of the momentum is completely localised at the origin. 
		By decreasing the value of $\nu$, the value of the metric component at $r=0$ is decreasing until it vanishes at $\nu =0$. 
		At the same time, a new maximum appears, signalling that the source of momentum is a null wave orbiting the centre.  
		In both plots we took $Q_1 = Q_5 = 10^5$ and $Q_P = 2\times 10^3$.
	}
	\label{fig:Plotgvv2}
\end{figure}

Varying $Q_A$ has an interesting effect on the geometry, and to describe this we  introduce a parameter  $\nu$, defined by:
\begin{align}
	\label{eq:nuDef}
	Q_A ~\equiv~ \sqrt{Q_5\,Q_P (1- \nu)}\,, \qquad\qquad  0 \leq \nu\leq 1\,.
\end{align} 
We then track the profiles $g_{vv}(r)$  and $g_{yy}(r)$ for various values of $\nu$ (see figure~\ref{fig:Plotgvv2}).   The former shows the profile of the momentum source, and the latter  shows how the size of the $S^1(y)$ depends on $r$.

The $g_{vv}(r)$ profile vanishes as $\sim r^{-2}$ at infinity because it asymptotes to the electric potential of the momentum charge.  At $r=0$ one has:
\begin{align}
	g_{vv}(r=0)~=~ 2\, \frac{ Q_P}{\sqrt{Q_1\,Q_5}} \, \nu  \,.
\end{align} 
The dependence on $\nu$ reflects the impact of the vector multiplet on the near horizon region.  

There are three regimes that emerge from the $g_{yy}$ profiles:  (i) A plateau with $g_{yy}$ taking a positive constant value near $r=0$, corresponding to an AdS$_2$ $\times S^1$ throat of a black hole,  (ii) Quadratic growth in $r$ corresponding to an AdS$_3$ region, and (iii) Another plateau with $g_{yy}$ limiting to a positive constant value as $r \to \infty$, corresponding to  flat space.

For $\nu >0$, there is a black hole and one sees all three regimes in the  $g_{yy}$ profile.  For $\nu =0$ there is a singular core with zero horizon area and there is no AdS$_2$ $\times S^1$ region and   the size of the $y$-circle vanishes at $r=0$. 
Despite having three asymptotic charges, the solution  $\nu =0$  behaves like a two-charge solution near the brane source.
In fact, it is this precise cancellation of charges that is responsible for the two-charge behaviour of the F1-NS5-P solution with local D0-D4 charge densities.

By changing the value of $\nu$ we can thus relate a three-charge black hole with a finite-size horizon to a solution where the horizon area vanishes, all the while keeping the momentum charge fixed.  Indeed, the asymptotic behaviour of the metric stays the same as one varies $\nu$.  What does change is the location of the momentum source.
For $\nu =1$, it is localised at $r=0$, the location of the branes. 
But as $\nu$ is decreased, and the effect of the vector multiplet increases, the momentum source gets spread outward and eventually becomes a null wave at $\nu =0$ where there is no momentum source at the origin. 

\subsection{F1-NS5-P supertube with D0-D4 charges}

By working directly in the F1-NS5-P frame, we can obtain much more general brane charge distributions.
As an example, we can consider an F1-NS5-P solution with dipolar D0-D4 charges with an arbitrary distribution of F1-NS5 branes in the base space.
In essence, we can combine the solutions of section~\ref{ssec:F1P} and section~\ref{ssec:NS5P} with a generic profile, $g_m(v)$, in the flat four-dimensional space.
In fact, one can consider many parallel strands  so that the harmonic functions take the form 
\begin{align}
	\label{eq:SupertubeHarm}
	Z_1 ~=~ c_1 +  \sum_{p=1}^n \frac{Q_{1p}} {\left|x_m - g_m(v) - a_m^{(p)}\right|^2}\,,\qquad
	Z_2 ~=~ c_2 + \sum_{p=1}^n \frac{Q_{5p}}{\left|x_m - g_m(v) - a_m^{(p)}\right|^2}\,,
\end{align}
where $Q_{1p}$ and $Q_{5p}$ are supergravity charges corresponding to each strand of the F1 and NS5 branes wrapping the $v$ circle which are located at $x_m = g_m(v) + a_m^{(p)}$ in the base space with $a_m^{(p)}$ denoting arbitrary constant shifts.
We have also introduced general constants $c_{1,2}$ which determine the asymptotic behaviour of the solution. By setting $c_1 = c_2 = 1$ one obtains the standard asymptotically-flat solution.
The rest of the Ansatz quantities are given by
\begin{gather}
	Z_A =- \dot F(v)(Z_2-c_2)\,, \qquad \cF=  -\sum_{m=1}^4\,\dot g_m(v)^2\,Z_1\,Z_2\,  -\dot F(v)^2\,\left(Z_2 - c_2\right)\,,\nonumber\\*
	h_{mn}=\delta_{mn}\,,\qquad \beta = \psi=\tilde A = 0\,,\qquad \omega = Z_1\,Z_2\,\dot g_m(v)dx^m\,,	\label{eq:SupertubeAnsatz}\\*
	\Theta^1 = \left(1+ *_4\right)d_4\left(Z_2\,\dot g_m(v)\,dx^m\right)\,,\qquad \Theta^2 = \left(1+ *_4\right)d_4\left(Z_1\,\dot g_m(v)dx^m\right)\,.\nonumber
\end{gather}

\section{Summary}
\label{sec:Summary}

In this section we make a ``stand-alone'' summary by pulling together the full Ansatz for the bosonic fields and collect all relevant BPS equations.   
Since it is likely to be most useful in applications, we also present the simplified system in which the base space metric is taken to be $v$-independent and hyper-K\"ahler and the one-form $\beta$ is $v$-independent.

\subsection{Supersymmetric Ansatz}

The bosonic content of the theory is the metric $g_{MN}$, a two-form gauge field $B$, a one-form $A$ and a dilaton $\phi$. 
In a supersymmetric solution, the most general form of the metric has the form:
\begin{align}
	ds_6^2  ~=~-\frac{2}{\sqrt{Z_1\, Z_2}}\,(d v+\beta)\,\left[d u+\omega + \frac{1}{2}\,\left(\cF + \frac{Z_A^2}{Z_2}\right)(d v+\beta)\right]~+~\sqrt{Z_1\, Z_2}\,d s^2_4(\cB)\,,\label{eq:SumMetAns2}
\end{align}
where the metric for the four-dimensional base space, $\cB$, will be denoted by:
\begin{align}
	\label{eq:SumBaseMet}
	d s^2_4(\cB) ~&=~ h_{mn}\, dx^m\, dx^n\,.
\end{align}
The fields $\beta$ and $\omega$ are one-forms on $\cB$, while $Z_1, Z_2, Z_A$ and $\cF$ are functions.

Note that because of the shift (\ref{eq:cFShift}), this metric is slightly different from the usual form of the BPS metric \cite{Gutowski:2003rg,Cariglia:2004kk}  that appears in the literature.

The dilaton and the gauge potentials can be decomposed as:
\begin{subequations}
	\label{eq:SumBosFields}
	\begin{align}
		e^{2\phi}~ & = ~\frac{Z_2}{Z_1}\,,\\
		A ~& =~ \frac{Z_A}{Z_2}\,(dv + \beta) - \tA\,,\\
		B ~& =~ -\frac{1}{Z_2}\, (du +\omega)\wedge (dv + \beta)+ \left[ a_2 - \frac{Z_A}{Z_2} \, \tA\right] \wedge (dv + \beta) + \gamma_1\,,
	\end{align}
\end{subequations}
where $\tA$, $a_2$ are one-forms and $\gamma_1$ is a two-form on $\cB$.

It is also convenient to introduce an auxilliary potential for the Hodge dual of the three-form field strength:
\begin{align}
	\widetilde B ~& =~ -\frac{1}{Z_1}\, (du +\omega)\wedge (dv + \beta)+ a_1 \wedge (dv + \beta) + \gamma_2\,.
\end{align}

The fields are thus: Scalar functions: $Z_1$, $Z_2$, $Z_A$ and $\cF$; one-forms on $\cB$: $\beta$, $\omega$, $\tA$, $a_1$, and $a_2$; and two-forms on $\cB$:  $\gamma_1$ and $\gamma_2$. 
Supersymmetry constrains all the fields to be independent of the coordinate $u$, but are allowed to depend on $v$ and the base coordinates, $x^m$.

The gauge-invariant field strengths are defined as
\begin{align}
	\label{eq:FieStrDef}
	F~=~ dA\,, \qquad G ~=~ dB + F\wedge A\,, \qquad 	-e^{2\phi} *_6 G~ =~ d\widetilde B\,,
\end{align}
and are given by
\begin{subequations}
	\label{eq:SumGaugeStrengths}
	\begin{align}
		F ~&  =~  (dv+ \beta) \wedge \omega_F + \tF\,,\\
		G ~&=~ d\left[ -\frac{1}{Z_2}\, (du +\omega)\wedge (dv + \beta)\right] + \widehat G_2\,,\\
		-e^{2\phi} *_6 G ~& = ~  d\left[ -\frac{1}{Z_1}\, (du +\omega)\wedge (dv + \beta)\right] + \widehat G_1\,,
	\end{align}
\end{subequations}
where 
\begin{subequations}
	\label{eq:SumGhats}
	\begin{align}
		\widehat G_1 ~& =~ *_4 \left(\cD Z_2 + \dot \beta Z_2 \right) + (dv+ \beta)\wedge \Theta^1\,,\\
		\widehat G_2 ~&=~*_4 \left(\cD Z_1 + \dot \beta Z_1 \right) + (dv+ \beta)\wedge\left( \Theta^2+ \frac{Z_A^2}{Z_2^2}\, \cD\beta -2 \frac{Z_A}{Z_2}\, \cD \tA\right)\,.
	\end{align}
\end{subequations}
These decompositions define the one-form, $\omega_F$, and the two-forms, $\Theta^1$ and $\Theta^2$, on $\cB$.

Comparison between the expressions for the gauge potentials \eqref{eq:SumBosFields} and the field strengths lead to the following expressions.
The components of the two-form field strength are given by
\begin{align}
	\label{eq:U1Comp}
	\omega_F =-\dot \tA + \frac{Z_A}{Z_2} \, \dot \beta - \cD \left(\frac{Z_A}{Z_2}\right)\,,\qquad 
	\tF =- \cD \tA + \frac{Z_A}{Z_2}\, \, \cD \beta\,,
\end{align}
while for the three-form gauge field and its dual we find
\begin{subequations}
	\begin{gather}
		\Theta^1~ =~ \cD a_1 - \dot \beta\wedge a_1 + \dot \gamma_2\,,\\ 
		*_4\left(\cD Z_2 + Z_2\,\dot \beta\right) ~=~ \cD \gamma_2 - a_1 \wedge \cD\beta\,,
	\end{gather}
\end{subequations}
and
\begin{subequations}
	\begin{gather}
		\Theta^2 ~=~ \cD a_2 - \dot \beta \wedge a_2+ \dot \gamma_1 + \dot \tA \wedge \tA\,,\\
		*_4\left(\cD Z_1 + Z_1\,\dot \beta\right) ~=~ \cD \gamma_1+ \cD \tA \wedge \tA  - a_2 \wedge \cD\beta\,.
	\end{gather}
\end{subequations}

There are still some unfixed gauge invariances in this system and these may be used to make convenient choices for $Z_A$, $\tA$, $a_1$,  $a_2$, $\gamma_1$ and $\gamma_2$.

\subsection{BPS equations}

The BPS equations can be organised into a triangular system with several layers.  The ``zeroth layer'' is the only non-linear system and it determines the base geometry on $\cB$, and  the fibration defined by $\beta$.  The remaining BPS equations are all linear  within the background defined in the zeroth layer.  For each layer, the linear equations are sourced quadratically by fields determined in the previous layers.

\subsubsection{The zeroth layer}

The base space, $\cB$, must  be almost hyper-K\"ahler\footnote{As far as we can tell, this is not a structure that has been extensively studied by mathematicians and so rather little is known about determining such geometries when they depend non-trivially on $v$.  If the background is $v$-independent then it is simply  hyper-K\"ahler, and there is a vast literature on such geometries. It is also important to remember that this base geometry can be ambi-polar  \cite{Giusto:2004kj,Bena:2005va,Berglund:2005vb, Bena:2007kg,Warner:2019jll}. }  and is determined by a set of three two-forms, $J^A$, and a fibration vector, $\beta$, that satisfy:

\begin{subequations}
	\label{eq:SumLay0J1}
	\begin{align}
		*_4 J^A ~&=~ - J^A\,,\\
		(J^A)^m{}_{n}(J^B)^n{}_{p}~&=~ \epsilon^{ABC}\,(J^C)^m{}_{p}-\delta^{AB}\,\delta^m_p\,,\\
		J^A \wedge J^B ~&=~ - 2\,\delta^{AB}\, {\rm vol}_4\,,\\
		d_4 J^A ~&=~ \pd_v\left(\beta \wedge J^A\right)\,,
	\end{align}
\end{subequations}
and
\begin{align}
	\label{eq:Dbetaeq1}
	*_4 \cD\beta~ =~ \cD\beta\,,
\end{align}
where the differential operator $\cD$ is defined in \eqref{eq:DefcalD} and Hodge duality is defined in (\ref{eq:HDDef}).

To write the BPS equations one needs to define an anti self-dual two-form:
\begin{align}
	\psi ~\equiv~  \frac{1}{8} \,\epsilon_{ABC}\, (J^A)^{mn}\, (\dot J^B)_{mn} \,J^C\,,
\end{align}

Note that if the background is $v$-independent then the metric must be hyper-K\"ahler, the self-duality equation, (\ref{eq:Dbetaeq1}), for $\beta$ is linear and $\psi \equiv 0$.

\subsubsection{The first layer}

The first step is to  determine the pair $(Z_2, \Theta^1)$
\begin{subequations}
	\label{eq:Lay1SubLay11}
	\begin{gather}
		*_4 \Theta^1   ~=~  \Theta^1  -  2 \,  Z_2\,  \psi\,,\\
		\cD *_4 \left[ \cD Z_2 + Z_2 \,\dot \beta\right] ~+~ \Theta^1 \wedge \cD \beta ~=~ 0\,,\\
		d_4 \Theta^1 ~=~ \pd_v \left[ \beta \wedge \Theta^1 +  *_4 \left(\cD Z_2 + Z_2\, \dot \beta\right)\right]\,.
	\end{gather}
\end{subequations}
%

\subsubsection{The second layer}

These equations determine the fields in the vector multiplet, and can be written in two equivalent ways.  In terms of field strength one has:
\begin{subequations}
	\begin{gather}
		*_4 \tF~= ~\tF\,,\\
		2\, \cD Z_2 \wedge *_4\omega_F ~+~ Z_2\, \cD *_4 \omega_F~=~ - \tF \wedge \Theta^1\,,
	\end{gather}
\end{subequations}
or, in terms of components of the potential $A$, one has:
\begin{subequations}
	\label{eq:Lay1Sub21}
	\begin{gather}
		*_4 \cD \tA   ~=~ \cD \tA\,,\\
		*_4 \cD Z_A \wedge \dot \beta \,+\,2\, \cD Z_2 \wedge *_4 \dot \tA \,+\, Z_2\,  \cD * \dot\tA \,+\, \cD *_4 \cD Z_A~=~  -\cD \tA \wedge \Theta^1\,,
	\end{gather}
\end{subequations}
%

\subsubsection{The third layer}

These equations determine  $(Z_1, \Theta^2)$:
\begin{subequations}
	\label{eq:Lay1SubLay22}
	\begin{gather}
		*_4  \Theta^2   ~=~  \Theta^2  -  2 \,  Z_1\,  \psi\,,\\
		\cD *_4 \left[ \cD Z_1 + Z_1 \,\dot \beta\right] ~+~ \Theta^2 \wedge \cD \beta ~=~ \cD \tA \wedge \cD \tA\,,\\
		d_4 \Theta^2 ~= ~\pd_v \left[ \beta \wedge\Theta^2 +  *_4 \left(\cD Z_1 + Z_1\, \dot \beta\right)\right]~-~ 2 \dot \tA  \wedge \cD \tA \,.
	\end{gather}
\end{subequations}
%

\subsubsection{The fourth layer}

This  final layer determines the one-form $\omega$, which encodes the information about the angular momentum in the system, and the scalar function $\cF$ whose asymptotic  behaviour determines the momentum.  
We have:
\begin{align}
	\label{eq:SumLay2eq11}
	\cD \omega + *_4 \cD \omega +  \cF \,\cD\beta ~=~  Z_1\, \Theta^1 + Z_2\, \Theta^2- 2\, Z_A\, \cD\tA - 2\, Z_1\,Z_2\,\psi \,,
\end{align}
and 
\begin{align}
	\label{eq:SumLay2eq2}
	&*_4 \cD *_4   L+ 2 \,\dot \beta^m \,  L_m \nonumber\\
	&=- \frac12 *_4 \left( \Theta^1 - Z_2 \, \psi\right)\wedge \left(\Theta^2 - Z_1 \, \psi\right) +  \ddot Z_1 \, Z_2 + \dot Z_1\, \dot Z_2 + Z_1 \, \ddot Z_2 \nonumber \\
	&\quad +\frac12 Z_1 \,Z_2 *_4  \psi\wedge  \psi +  *_4  \psi \wedge \cD \omega- \frac14 Z_1\, Z_2\, \dot h^{mn}\,\dot  h_{mn} + \frac12 \pd_v \left[Z_1 \, Z_2 \,h^{mn}\dot h_{mn} \right]\nonumber\\*
	& \quad + Z_2 \,\dot \tA^2 +  \dot\tA_m \left[\cD Z_A\right]^m\,,
\end{align}
where we defined a one-form
\begin{align}
	L \equiv  \dot \omega + \frac{\cF}{2} \,\dot \beta- \frac12\, \cD  \cF+ Z_A\, \dot\tA\,.
\end{align}
and the indices are raised and lowered with respect to the four-dimensional base metric $h_{mn}$.

\subsection{Equations with a flat base}
\label{app:FlatBase}

Since very little is known about generic  almost hyper-K\"ahler manifolds, the first step in practice is to specialize to backgrounds in which the base metric and fibration vector, $\beta$, are $v$-independent.  However all other fields are still allowed to depend on $v$.   This results in a very significant simplification of the system.  As we remarked above, the base geometry is then (ambi-polar) hyper-K\"ahler, and the  equation for $\beta$, (\ref{eq:Dbetaeq1}) becomes linear self-duality:
\begin{align}
	\label{eq:ADbetaeqvind}
	*_4 d_4\beta ~=~ d_4\beta\,.
\end{align}
The subsequent layers of equations are now:
\begin{subequations}
	\label{eq:ALay1SubLay1vind}
	\begin{gather}
		*_4 \Theta^1   ~= ~ \Theta^1  \,,\\
		\cD *_4 \cD Z_2  + \Theta^1 \wedge d_4 \beta ~=~ 0\,,\\
		\cD \Theta^1 ~=~ *_4 \cD \dot Z_2\,;
	\end{gather}
\end{subequations}
with the second layer: 
\begin{subequations}
	\label{eq:ALay1Sub2vind}
	\begin{gather}
		*_4 \cD \tA   ~= ~\cD \tA\,,\\
		2 \cD Z_2 \wedge *_4 \dot \tA + Z_2  \cD * \dot\tA + \cD *_4 \cD Z_A~=~  -\cD \tA \wedge \Theta^1\,;
	\end{gather}
\end{subequations}
and the third layer: 
\begin{subequations}
	\label{eq:ALay1SubLay22vind}
	\begin{gather}
		*_4  \Theta^2   ~=~  \Theta^2 \,,\\
		\cD *_4 \cD Z_1  ~+ ~\Theta^2 \wedge d_4 \beta ~=~ \cD \tA \wedge \cD \tA\,,\\
		\cD \Theta^2 ~=~ *_4 \cD \dot Z_1~-~ 2 \dot \tA  \wedge \cD \tA \,;
	\end{gather}
\end{subequations}
and the fourth layer: 
\begin{align}
	\cD \omega + *_4 \cD \omega ~=~  Z_1\, \Theta^1 + Z_2\, \Theta^2 -  \cF \,d_4\beta- 2 Z_A\, \cD\tA \,,
\end{align}
\begin{align}
	&*_4 \cD *_4   L ~=~- \frac12 *_4 \Theta^1 \wedge \Theta^2  +  \ddot Z_1 \, Z_2 + \dot Z_1\, \dot Z_2 + Z_1 \, \ddot Z_2  + Z_2 \,\dot \tA^2 +  \dot\tA_m \left[\cD Z_A\right]^m\,,
\end{align}
where now
\begin{align}
	  L ~=~  \dot \omega - \frac12\, \cD  \cF+ Z_A\, \dot\tA\,.
\end{align}
%

\section{Final comments}
\label{sec:Discussion}

We have shown that the BPS equations for $\cN = (1,0)$ supergravity in six-dimensions with a U(1) vector multiplet are linear, with the triangular layered structure encountered in other simpler theories    \cite{Bena:2004de,Bena:2005va, Bena:2007kg,Bena:2011dd,Giusto:2013rxa}.  We have also presented some simple, but illustrative solutions of this BPS system.

We expect that this work will open up the construction and exploration of new classes of superstrata much as the linearization of \cite{Bena:2011dd,Giusto:2013rxa} opened up the construction of D1-D5-P, or IIB, superstrata.  It is  important to note that ``IIA superstrata'' obtained from our system will be intrinsically different from their IIB, or D1-D5-P, counterparts.  In particular, the  IIA superstrata cannot be transformed directly into IIB superstrata because this would involve either performing a T-duality along a direction of the $\IT^4$ thereby generically breaking the assumed $\IT^4$-invariance, or taking a T-dual along the $y$-direction which in general is not a symmetry direction of  superstrata.  Indeed, the latter transformation would result in intrinsically-stringy states that are not described by supergravity.

One might imagine that a suitably inventive chain of dualities involving all the compact directions might map IIA superstrata onto IIB superstrata, but if that were possible then one would be able to resolve the troublesome degenerate black-hole limit of IIB superstrata, discussed in \cite{Bena:2022sge}, entirely within the IIB frame.  This has not been achieved and became the motivating force behind   \cite{Bena:2022sge} to  go to the IIA frame,  where the resolution is based on the ``new momentum carriers.'' Thus the supergravity theory analysed in this paper is exploring a {\it new} branch of microstate geometries.

Our examples, and the results of  \cite{Bena:2022sge}, illustrate some of the physics of this new branch, especially the fact that the momentum charge of a three-charge geometry can be diffused into the vector-multiplet excitations, leaving an effective two-charge core to the geometry.  We expect that this is also how the new momentum carriers remove the horizon in the degenerate limit of IIB superstrata. On the other hand, the examples presented here were obtained primarily through chains of dualities, and did not really exploit the linearization of the system. Moreover, these solutions, despite having vanishing horizon area, are   still singular.  

We expect that there are  much richer families of IIA superstrata with completely smooth geometries and supported by non-trivial oscillating fluxes, much like the IIB superstrata.  There are, however, some significant differences with IIB superstrata.  First, the non-trivial fluxes of the IIB geometries are all dual to 3-cycles, whereas in the IIA system, the momentum carriers are described by a 2-form field strength and this suggests one will need non-trivial 2-cycles. This indicates that the starting point of smooth IIA superstrata might be the 2-centered, ambi-polar GH metric rather than a supertube in flat space.  The ``degree of difficulty'' here is no more than the IIB superstratum because the wave equation is still separable in spherical bipolar coordinates on two-centered base spaces. Moreover, it is already known how to construct fluctuating 2-form fluxes on such spaces \cite{Bena:2017geu,Tyukov:2018ypq,Walker:2019ntz}.

The next difference is the separation of layers in the IIA system compared to the IIB system.   At a technical level this means that the smoothness constraints and the ``coiffuring'' \cite{Bena:2013ora,Bena:2014rea,Bena:2015bea} will be somewhat different in the IIA system.  More specifically, coiffuring becomes important when the linear systems are sourced by squares of fluctuating quantities.  For IIB superstrata, this only happens in the final layer that determines $\cF$ and $\omega$.  However, for non-trivial momentum carriers in our system, it is the vector field $A$ that fluctuates and this field quadratically sources  $(Z_1, \Theta^2)$, as one sees in  (\ref{eq:Lay1SubLay22}).  Thus we anticipate that coiffuring will become important in this earlier layer of the system. 

At a more physical level, this observation means  that the branes and momentum excitations are interacting in a different manner and this could lead to new physics.   Indeed, the construction of non-trivial IIA superstrata will  shed light on the localisation of momentum charge within the solution constructed in \cite{Bena:2022sge}.  In our simple toy examples we saw that there was a competition between the momentum potential, $\cF$, and the electrostatic potential, $Z_A$, that leads to a spreading of the momentum charge distribution in the core of the solution.  It would be extremely interesting to see how that plays out in asymptotically-flat IIA superstrata and whether smoothness (or ``coiffuring'') limits the momentum charge distribution. 

Another important line for future work is to construct the embedding of the six-dimensional theory discussed here into M-theory.  We have motivated our choice of the six-dimensional theory through the IIA description of the momentum carriers, and we have understood the role of the extra vector multiplet as coming from the compactification of the $x^{10}$-circle in M-theory.  Ideally we would like to know exactly how to uplift our solutions to a full M-theory solution.  Such knowledge has proven invaluable in understanding and developing IIB superstrata and we would expect similar insights would come from a complete M-theory description of IIA superstrata.  

Indeed, such M-theory uplifts would lead to some potentially interesting generalizations.  For example, one could try to create a deeper democracy between the $x^{10}$ circle and the $y$ circle, and perhaps seek out microstate geometries that depend upon $x^{10}$ as well as $v$.   The most general such system will involve  $x^{10}$ and $v$ fibration vectors that depend on both coordinates and thus the non-linear ``zeroth layer" of the BPS system could be very challenging.   On the other hand, there might be a very interesting, and manageable subsystem in which $\beta$ is independent of $v$ but depends on $x^{10}$, while the fibration vector of $x^{10}$ is independent of $x^{10}$ but depends on $v$.

This raises a further and broader question about linearizing BPS systems.  When originally discovered, the linearization of the BPS equations of five-dimensional $\Neql{2}$ supergravity coupled to vector multiplets appeared to be an isolated miracle.  However, the subsequent generalization to six dimensions, and the work presented here suggests that the miracle might well extend to other systems.   An obvious next step, motivated by the potential M-theory generalizations listed above, would be to look at the BPS system of minimal seven dimensional supergravity, with the M-theory circle providing the uplift to seven dimensions.
 
 Finally, there is the much broader issue of incorporating new physics coming from the compactified dimensions of microstate geometries.  In the early days,  microstate geometries were restricted to five dimensions simply out of expediency.  It became evident that while such five-dimensional geometries might be able to capture the ground states of phases of the black-hole field theory \cite{Bena:2013dka},  one would need to allow fluctuations that depend on the extra dimensions if one wanted to  capture a significant amount of the microstructure of black holes.  This was the driving force behind the development of superstrata. One of the hopes was that going to six dimensions might be sufficient to capture the entropy growth.  The idea was that  the $y$-circle represents the world volume of the CFT and the operators with space-time polarizations are sufficient to capture the entropy growth  \cite{Bena:2014qxa}.  If these states can be seen in supergravity then six dimensions would indeed be sufficient, and  adding more dimensions would simply amount to adding more transverse polarizations thereby increasing the central charge to its correct value.  
 
 It is now evident that this perspective was too naive.  First, the superstrata constructed to date do not seem to be able resolve any details of the twisted sectors of the CFT.  There are also strong arguments that the full black-hole microstructure must involve non-trivial physics in the internal dimensions.  For example, W-branes \cite{Martinec:2015pfa,Martinec:2019wzw,Martinec:2020gkv} and other Higgs branch condensates may well require   such non-trivial dynamics. The results in \cite{Bena:2022sge} and the techniques  developed here suggest that dynamics of the internal dimensions are also essential to resolving degenerate corners of the moduli space of IIB superstrata.  There has also been some remarkable progress in constructing non-extremal microstate geometries  using the  charged Weyl formalism to solve the full bosonic equations of motion  \cite{Bah:2020pdz, Bah:2021owp, Bah:2021rki,Heidmann:2021cms, Bah:2022yji}.  These new non-BPS microstate geometries make essential use of topological transitions and ``bubbles'' on the compactification manifolds.  
 
 It is now clear that microstate geometries can have extremely non-trivial  dynamics in the compactification directions and that understanding this will be essential to a full understanding of black-hole microstructure.  This paper represents a modest step in developing some of the essential tools to achieving this.

\section*{Acknowledgements}

%
We would like to thank  our collaborators on \cite{Bena:2022sge}, and especially Iosif Bena for discussions and invaluable input on this project. We would also like to thank  Anthony Houppe, Bogdan Ganchev, Bin Guo, and Daniel R. Mayerson for interesting discussions.
This work  is supported by the ERC Grant 787320 - QBH Structure.
The work of NPW is also supported, in part, by the DOE grant DE-SC0011687.

\appendix

\section{BPS equations in gauge invariant form}
\label{app:NoShift}

In the main text we  presented the supersymmetric Ansatz  that leads to a natural linear structure.
While the analysis was driven by the desire to lay out this linear structure, each component of the Ansatz corresponds to a particular source in the higher-dimensional  supergravity. 

It is possible that the form of the equations presented in the main text is not well-suited to all applications.
To that end, we use this Appendix to present the results of this paper in a slightly different manner, which are more in line with the analysis of 
\cite{Bena:2011dd, Giusto:2013rxa}. 
While this second presentation obscures the direct  identification between Ansatz quantities and objects in higher dimensional theories of gravity, it emphasises quantities that are invariant under the U(1) gauge transformation. 

Since the linearization of the supersymmetric Ansatz and BPS equations follows the exact same reasoning as in the main text, we simply state the Ansatz and the associated BPS equations, highlighting any relevant differences.
Our aim is to make this appendix reasonably self-contained at a cost of repetition of some equations from the main text.
The key results are the same: The BPS equations organise themselves into several layers of linear differential equations, which have additional complexity due to the addition of the Maxwell fields, compared to their D1-D5-P counterparts.

\subsection{Ansatz}

The six-dimensional metric has the more standard BPS form \cite{Gutowski:2003rg,Cariglia:2004kk}:
\begin{subequations}
	\begin{align}
		ds_6^2  ~&=~-\frac{2}{\sqrt{Z_1\, Z_2}}(d v+\beta)\Big[d u+\omega + \frac{\widetilde{ \mathcal{F}}}{2}(d v+\beta)\Big]~+~\sqrt{Z_1\, Z_2}\,h_{mn}\, dx^m\, dx^n\,,	\label{eq:AMetAns1}
	\end{align}
\end{subequations}
with other bosonic fields in the theory given by 
\begin{subequations}
	\label{eq:AFields}
	\begin{align}
		e^{2\phi} ~& =~ \frac{Z_2}{Z_1}\,,\\
		F ~&  =~  (dv+ \beta) \wedge \omega_F + \tF\,,\\
		G ~&=~ d\left[ -\frac{1}{Z_2}\, (du +\omega)\wedge (dv + \beta)\right] + \widehat G_2\,,
	\end{align}
\end{subequations}
and we also introduce
\begin{align}
	-e^{2\phi} *_6 G ~& = ~  d\left[ -\frac{1}{Z_1}\, (du +\omega)\wedge (dv + \beta)\right] + \widehat G_1\,.
\end{align}
The two-forms $\widehat G_{1,2}$ can be further decomposed as
\begin{subequations}
	\begin{align}
		\widehat G_1 ~& \equiv~ *_4 \left(\cD Z_2 + \dot \beta Z_2 \right) + (dv+ \beta)\wedge \Theta^1\,,\\
		\widehat G_2 ~& \equiv~ *_4 \left(\cD Z_1 + \dot \beta Z_1 \right) + (dv+ \beta)\wedge\widetilde\Theta^2\,.
	\end{align}
\end{subequations}
The gauge field strengths can be written in terms of lower-form potentials
\begin{align}
	\label{eq:AFieStrDef}
	F~=~ dA\,, \qquad G ~=~ dB + F\wedge A\,, \qquad 	-e^{2\phi} *_6 G ~=~ d\widetilde B\,,
\end{align}
which can be also decomposed as
\begin{subequations}
	\label{eq:APotentials}
	\begin{align}
		A ~& =~ \frac{Z_A}{Z_2}\,(dv + \beta) - \tA\,,\\*
		B ~& =~ -\frac{1}{Z_2}\, (du +\omega)\wedge (dv + \beta)+ \widetilde a_2 \wedge (dv + \beta) + \gamma_1\,,\\*
		\widetilde B ~& =~ -\frac{1}{Z_1}\, (du +\omega)\wedge (dv + \beta)+ a_1 \wedge (dv + \beta) + \gamma_2\,.
	\end{align}
\end{subequations}
Reconciling \eqref{eq:APotentials} with \eqref{eq:AFields} leads to the identifications:
\begin{align}
	\label{eq:AU1Comp}
	\omega_F =-\dot \tA + \frac{Z_A}{Z_2} \, \dot \beta - \cD \left(\frac{Z_A}{Z_2}\right)\,,\qquad 
	\tF =- \cD \tA + \frac{Z_A}{Z_2}\, \, \cD \beta\,,
\end{align}
for the U(1) gauge field strength, with  
\begin{subequations}
	\begin{gather}
		\Theta^1 ~=~ \cD a_1 - \dot \beta\wedge a_1 + \dot \gamma_2\,,\\ 
		*_4\left(\cD Z_2 + Z_2\,\dot \beta\right) ~=~ \cD \gamma_2 - a_1 \wedge \cD\beta\,,
	\end{gather}
\end{subequations}
and
\begin{subequations}
	\label{eq:ACond2}
	\begin{gather}
		\widetilde\Theta^2 ~=~ \cD \widetilde a_2  - \dot \beta \wedge  \widetilde a_2 + \dot \gamma_1 + \tA \wedge \omega_F + \frac{Z_A}{Z_2}\, \tF  \\
		*_4\left(\cD Z_1 + Z_1\,\dot \beta\right) ~=~ \cD \gamma_1  - \widetilde a_2  \wedge \cD\beta- \tF\wedge \tA\,,
	\end{gather}
\end{subequations}
for the three-form gauge field strengths.
We note that under the U(1) gauge transformations \eqref{eq:U1Gauge} and \eqref{eq:U1GaugeB} the quantities $\widetilde a_2$ and $\gamma_1$ change as 
\begin{align}
	\label{eq:AU1GaugeBComp}
	\widetilde a_2 \to \widetilde a_2 + \frac{Z_A}{Z_2}\, \cD \lambda + \dot \lambda\,\tA\,, \qquad \gamma_1 \to \gamma_1 + \tA\wedge\cD\lambda\,,
\end{align}
while the right hand sides of \eqref{eq:ACond2}, and thus the components of $G$, are invariant. 
Hence $\widetilde\Theta^2$ is gauge invariant, unlike in the main text (see \eqref{eq:cFTh2GaugeTrans}). 
Similarly, due to the absence of explicit $Z_A$ terms in the metric, $\widetilde \cF$ is also invariant under this gauge transformation.

\subsection{BPS equations}

Once again, the BPS equations can be organised into several layers. 
Since most of them are exactly the same as in the main text, we simply state them without additional explanation.  (See in Section~\ref{sec:Summary} for more details.)
When using the fields $\widetilde \Theta^2$ and $\widetilde \cF$ it is more convenient to work in the manifestly U(1) gauge invariant form. 

\subsubsection{The zeroth layer}

The zeroth layer equations are unchanged and are given by the equations for the complex structures on the base space
\begin{subequations}
	\label{eq:ALay0J1}
	\begin{align}
		*_4 J^A ~&=~ - J^A\,,\\*
		(J^A)^m{}_{n}(J^B)^n{}_{p}~&=~ \epsilon^{ABC}(J^C)^m{}_{p}-\delta^{AB}\delta^m_p\,,\\*
		J^A \wedge J^B ~&=~ - 2\delta^{AB}\, {\rm vol}_4\,,\\*
		d_4 J^A ~&=~ \pd_v\left(\beta \wedge J^A\right)\,,
	\end{align}
\end{subequations}
and the self-duality condition for the one-form $\beta$
\begin{align}
	\label{eq:ADbetaeq1}
	*_4 \cD\beta ~=~ \cD\beta\,.
\end{align}

\subsubsection{The first layer}
%
The set of equations for $(Z_2, \Theta^1)$ is unchanged:
\begin{subequations}
	\label{eq:ALay1SubLay11}
	\begin{gather}
		*_4 \Theta^1   ~=~  \Theta^1  -  2 \,  Z_2\,  \psi\,,\\
		\cD *_4 \left[ \cD Z_2 + Z_2 \,\dot \beta\right] + \Theta^1 \wedge \cD \beta ~=~ 0\,,\\
		d_4 \Theta^1 ~=~ \pd_v \left[ \beta \wedge \Theta^1 +  *_4 \left(\cD Z_2 + Z_2\, \dot \beta\right)\right]\,,
	\end{gather}
\end{subequations}
where 
\begin{align}
	\psi ~=~  \frac{1}{8}\, \epsilon_{ABC} \,(J^A)^{mn} \,(\dot J^B)_{mn}\, J^C\,,
\end{align}
is a two-form that  captures the anti self-dual part of $\Theta^1$.

\subsubsection{The second layer}

The equations for the second layer, determining the components of the U(1) gauge field, can be expressed in a gauge-invariant form:
\begin{subequations}
	\begin{gather}
		*_4 \tF~=~ \tF\,,\\
		2\, \cD Z_2 \wedge *_4\omega_F \,+\, Z_2\, \cD *_4 \omega_F~=~ - \tF \wedge \Theta^1\,.
	\end{gather}
\end{subequations}
%

\subsubsection{The third layer}
The third layer, which determines, $(Z_1, \widetilde \Theta^2)$ is now:
\begin{subequations}
	\label{eq:ALay1SubLay21}
	\begin{gather}
		*_4 \widetilde\Theta^2   ~=~  \widetilde\Theta^2  -  2 \,  Z_1\,  \psi\,,\\
		\cD *_4 \left[ \cD Z_1 + Z_1 \,\dot \beta\right] + \widetilde\Theta^2 \wedge \cD \beta ~=~  \tF\wedge \tF \,,\\
		d_4 \widetilde\Theta^2 ~=~ \pd_v \left[ \beta \wedge \widetilde\Theta^2 +  *_4 \left(\cD Z_1 + Z_1\, \dot \beta\right)\right]- 2\, \omega_F \wedge \tF\,,
	\end{gather}
\end{subequations}
with the vector multiplet contributions appearing in the quadratic combinations of $\omega_F$ and $\tF$.

\subsubsection{The fourth layer}

The equations in the last layer are also slightly modified:
The first equation now reads
\begin{align}
	\label{eq:ALay2eq11}
	\cD \omega + *_4 \cD \omega + \widetilde\cF \,\cD \beta~=~   Z_1\, \Theta^1 + Z_2\,\widetilde\Theta^2 - 2 Z_1\,Z_2\,\psi \,,
\end{align}
and we note that there is no explicit contribution from the vector field.
However, its effect is implicitly hidden in $\widetilde \cF$.  Indeed,  this vector field makes an explicit appearance in the second equation through an $\omega_F^2$ in the source terms: 
\begin{align}
	\label{eq:ALay2eq21}
	&*_4 \cD *_4  \widetilde L+ 2 \,\dot \beta^m \,\widetilde  L_m  \nonumber\\*
	&= - \frac12 *_4 \left( \Theta^1 - Z_2 \, \psi\right)\wedge \left(\widetilde\Theta^2 - Z_1 \, \psi\right) +  \ddot Z_1 \, Z_2 + \dot Z_1\, \dot Z_2 + Z_1 \, \ddot Z_2  + Z_2\,\omega_F^2  \nonumber \\*
	&\quad +\frac12 Z_1 \,Z_2 *_4  \psi\wedge  \psi +  *_4  \psi \wedge \cD \omega- \frac14 Z_1\, Z_2\, \dot h^{mn}\,\dot  h_{mn} + \frac12 \pd_v \left[Z_1 \, Z_2 \,h^{mn}\dot h_{mn} \right]\,,
\end{align}
where we define:
\begin{align}
	\widetilde	L ~\equiv ~\dot \omega + \frac{\widetilde\cF}{2} \,\dot \beta- \frac12\, \cD\widetilde \cF\,,
\end{align}
and the square of $\omega_F$  is constructed using the four-dimensional base metric $h_{mn}$:
\begin{align}
	\omega_F^2~ =~ \left(\omega_F\right)_m\,\left(\omega_F\right)_n\, h^{mn}\,.
\end{align}

\bibliographystyle{JHEP}

\bibliography{lin}

\end{document}